\documentclass[aps,prd,amsmath,amssymb,11pt]{revtex4}
\usepackage{graphicx}
\usepackage{bm}
\def\journal#1#2#3#4{{#1} {\bf #2}, #3 (#4)}
\newcommand{\be}{\begin{equation}}
\newcommand{\ee}{\end{equation}}
\newcommand{\bea}{\begin{eqnarray}}
\newcommand{\eea}{\end{eqnarray}}
\newcommand{\hf}{\frac12}
\newcommand{\nn}{\nonumber\\}
\def\eq#1{(\ref{#1})}

\def\fd#1#2{\frac{\delta#1}{\delta#2}}

\def\la{\langle}
\def\ra{\rangle}

\def\mr#1{{\mathrm{#1}}}
\def\ord#1{{\cal O}(#1)}

\def\dt{{\Delta t}}
\def\hj{{\hat j}}
\def\hx{\hat x}
\def\hy{\hat y}
\def\hD{{\hat D}}
\def\hG{{\hat G}}
\def\ih{\frac{i}{\hbar}}

\def\sign{\mr{sign}}

\def\v#1{{\bm{#1}}}
\def\Tr{{\mr{Tr}}}

\begin{document}
\title{Spontaneous breakdown of the time reversal symmetry}
\author{Janos Polonyi}
\email{polonyi@iphc.cnrs.fr}
\affiliation{Strasbourg University, CNRS-IPHC, 23 rue du Loess, BP28 67037 Strasbourg Cedex 2, France}

\begin{abstract}
The role of the environment initial conditions in the breaking of the time reversal symmetry of effective theories and in generating the soft irreversibility is studied by the help of Closed Time Path formalism. The initial conditions break the time reversal symmetry of the solution of the equation of motion in a trivial manner. When open systems are considered then the initial conditions of the environment must be included in the effective dynamics. This is achieved by means of a generalized $\epsilon$-prescription where the non-uniform convergence of the limit $\epsilon\to0$  leaves behind a spontaneous breakdown of the time reversal symmetry.
\end{abstract}
\maketitle
\tableofcontents

\section{Introduction}

One usually observes a system in interaction with its environment. The laws, discovered in such a manner are highly complex, they describe an open, effective rather than conserved, closed dynamics. Some simplifications may take place when we bring the environment into the thermodynamical limit and the complicated, open features of the dynamics are reduced to irreversibility and dissipative forces. The emergence of the ensuing dynamical breakdown of the time reversal invariance is the subject of this work.

The number of the dynamical variables of a large environment is controlled by an artificially introduced quantity, the cutoff, and its removal defines the infinitely large environment. The cutoff is usually a dimensional number and is chosen to be far away from the scale of observations to make the predictions approximatively independent of its choice. One the one hand, this procedure makes our description natural in the sense that it eliminates an unreasonably large number in our expressions. But, on the other hand, the resulting infinite system is defined formally and it may possess surprising features if the removal of the cutoff takes place in a non-uniformly convergent manner. 

Our intuition is based on uniform convergence, and a point-wise convergence can produces phenomenas which are not localizable in finite space-time or energy-momentum regions. Multiple limits may develop a dependence on the order of their execution, multiple integrals may become dependent on the order of integration and the result of the execution of a limit and a derivative of a parametric integrals may depend of the order the limit and the derivative are taken. A well known family of non-uniformly converging integrals consists of the distributions, the generalized functions: The regulated distribution is a regular function which depends on a parameter $\epsilon$ and the limit, $\epsilon\to0$, is supposed to be carried out after the integration only. 

Non-uniform convergences appear usually through the amplification by a divergence, controlled by a cutoff. The infinity can hide at infinitely small or at infinitely large scales and we have three kinds of dimensional variables, length, time and mass hence there are six different types of infinity. The mass-dependence can usually by expressed by the help of wavelengths, the remaining four infinities being the following: (i) $L\to\infty$: The thermodynamical limit generates the spontaneously broken symmetries. (ii) $L\to0$: The infinitely short distance modes lead to finite, observable effects, called anomalies in quantum field theories. (iii) $T\to0$: The short time structure of the trajectory of a point particle in non-relativistic quantum mechanics reveals an $\ord\hbar$ effect, the fractal nature of the particle trajectories. (iv) $T\to\infty$: The infinitely slow modes generate high sensitivity on the initial or final conditions and determine the fate of the time reversal invariance. These phenomena are briefly outlined in Appendix \ref{nonunifap} for completeness, and we restrict our attention to the limit $T\to\infty$ in the rest of this paper.

The initial or final conditions, commonly called auxiliary conditions, are carefully separated from the equations of motion because we can freely adjust the former whereas the latter is given by Nature. Our starting point is that such a separation is not justified in the effective dynamics of the observed system, interacting with its environment, because the effective dynamics obviously depends on the environment initial conditions in a non-local manner in time. In other words, the environmental initial conditions are an integral part of the effective system dynamics, rather than being an independent input. Such a state of affairs can be seen easier when the effective dynamics is approximated by expanding its non-locality in the time derivative. The resulting effective equation of motion contains higher order derivatives and needs additional auxiliary conditions to provide a unique solution. Although such data should obviously come from the environment, this latter being unobserved, we do not possess this information. The solution, proposed below is to incorporate the environmental auxiliary conditions into the effective action itself.

Once the environment initial conditions are taken care in the effective action they influence the symmetry of the dynamics. Trace the fate of time related symmetries when the system is followed in the time interval $i_i\le t\le t_f$: The time translation invariance is recovered in the limit $t_i\to-\infty$ if the initial conditions correspond to a stationary state of motion. The state of the time reversal symmetry is more involved. The loss of the time reversal symmetry can be detected for finite $t_f$ by recording the motion of the system and checking whether the recording played forward and backward satisfies the same equation of motion. There is an important difference between the motions running in opposite time directions, namely the initial conditions: The initial conditions of the playback motion are the final conditions of the original one. This fact is crucial for us since the environment trajectory depends on the environment auxiliary conditions and it is just this dependence that breaks the time reversal invariance of the effective dynamics. The nontrivial question here is whether such a symmetry breaking survives the limits $t_i\to-\infty$ and $f_f\to\infty$.

We assume that the dynamics of the full system is closed and invariant under time reversal and show that the time reversal symmetry of the effective dynamics of the observed system is broken by a mechanism which is similar to that occurring in the spontaneous symmetry breakdown. In particular, one can separate the following three levels of the breakdown of the time reversal symmetry: (i) The auxiliary conditions of a closed system, presented separately from the equation of motion, leave the dynamics time reversal invariant and break the symmetry of the solutions only. This effect, not being part of the dynamical equations, is to be compared with the trivial, external symmetry breaking. (ii) The initial conditions of a closed system will be represented by a generalized $\epsilon$-prescription, namely by some infinitesimal terms in the action in such a manner that the time reversal transformation flips the sign of these terms, $\epsilon\to-\epsilon$. The time reversal invariance is broken by an infinitesimal, $\ord\epsilon$ symmetry breaking term in the action which generates a finite, $\ord{\epsilon^0}$ effect on the level of the solutions. This phenomenon is analogous to the spontaneous symmetry breaking except that it takes place even for a single degree of freedom and leaves the finite $\ord{\epsilon^0}$ part of the action untouched. (iii) By virtue of a symmetry breaking {\em within the environment} which is easy to excite, ie. has gapless spectrum the effective action contains finite $\ord{\epsilon^0}$ symmetry breaking terms which may generate soft irreversibility. The possible relation between irreversibility and dynamically broken symmetries has already been  noticed, namely that irreversible systems need an infinitely large environment with continuous spectrum \cite{prigogine,prigoginebb,kamenev,dynbr}. The goal of the present work is to explain in detail that a large environment can make the effective dynamics irreversible by the non-uniform convergence as the different regulators are removed.

The possible breakdown of the time reversal symmetry by the boundary conditions are detected by comparing two motions, one follows the real time evolution from the initial to the final boundary conditions and the initial and the final conditions are exchanged for the other. This makes the Closed Time Path (CTP) formalism, characterized by the redoubling of the degrees of freedom and evolving the doublers in opposite direction of the time, well suited to address this problem. This scheme is available both in the quantum and the classical level and has further important advantages, namely it provides a unique solution of the effective equation of motion with higher oder time derivatives, it allows the dynamical breakdown of the time reversal symmetry by treating the initial conditions as part of the dynamics and finally it supports dissipative forces which are local in time. This scheme has been introduced long time ago in quantum mechanics \cite{schw} and has already been used in different contexts, such as the relaxation in many-body systems \cite{kadanoffb}, perturbation expansion for retarded Green-functions \cite{keldysh}, manifestly time reversal invariant description of quantum mechanics \cite{aharonov,griffith,hartleg}, finite temperature effects in quantum field theory \cite{umezawa,niemisa,niemisb}, mixed state contributions to the density matrix by path integral \cite{feynman}, non-equilibrium processes \cite{calzetta,kamenev}, equations of motion for the expectation value of local operators \cite{jordan,ed}  and scattering processes with non-equilibrium final states \cite{scatt}. The distinguished feature of the CTP scheme, a reduplication of the degrees of freedom, fits so naturally to quantum mechanics that one wonders if such a modification is not implicitly present already in classical mechanics. The result of such an inquiry is the classical CTP scheme \cite{arrow,galley}, to be used below.

The spontaneous symmetry breaking takes place in systems with symmetric local dynamics. Hence the spontaneous breakdown of the time reversal symmetry is a nothing but a dynamical proposal to solve part of the time arrow problem, the emergence of an orientation for the  time in a time reversal invariant microscopic dynamics \cite{eddington,reichenbach,zehta,mackey,schulmanta,mersinih,halliwellpz,savitt}. The simplest and most natural explanation of the directed flow of time in macroscopic phenomena is the second law of thermodynamics, the non-decreasing nature of entropy for closed systems. The generalization of the thermodynamical entropy to statistical mechanics, the Boltzmann and the Gibbs entropies \cite{lieb}, together with the information theoretical approach \cite{jaynes}, allow us to extend the entropy beyond equilibrium states. The dissipated work, given in terms of the relative entropy of a driven system and its time reversed form \cite{jarzynski,crooks,kawai,roldan}, relates the entropy production and the thermodynamical time arrow in simple and convincing manner. The retardation of the electromagnetic radiation, described by the time reversal symmetric Maxwell equations, is the manifestation of the radiation time arrow. When a phenomenon crosses the border of the quantum and the classical regimes and a unique reality is formed then the irreversible collapse of the wave function indicates the presence of a quantum time arrow, too. These time arrows, generated by time reversal invariant dynamics, arise from a common origin, namely the unusually low entropy, ordered and cohered initial state of the Universe \cite{lebowitz,price}. In other words, the breakdown of the time reversal symmetry should be generated by the ultimate cosmological time arrow. This turn of the argument opens the Pandora's box and confronts us a completely new problem, namely there might be subsystems with different initial and final entropy, supporting different time arrows, in particular the time arrow should be opposite in the expanding and in the contracting phase of the Universe \cite{hawkingcta,page}. The low initial entropy of the universe is due to the homogeneity and isotropy \cite{penrose} and it is natural to suspect inflation, the driving force to reach a homogeneous and isotropic state, as the origin of the time arrow \cite{davies}. There are counter arguments \cite{pageinfl,holland} proposing a different scenario where different space-time regions support different time arrows. The eternal \cite{aguirre} and the spontaneous inflation \cite{carroll} may offer a more refined framework however this picture is presumably not the final one.

A tiny time arrow was ignored in the previous paragraph which arises from the CP violating weak interactions and turns into T-violation by assuming CPT symmetry. In view of the importance of the time arrow issue we need the demonstration of the explicit T-violation by the weak interactions, a problem which turned out to be surprising difficult \cite{sachs}. The simplest check, namely the measurement of a non-vanishing expectation value of a T-odd observable in a non-degenerate stationary state, is not feasible. The reason is that in the studies of the weak interactions we have access to scattering states only where the different initial and final state interactions generate T-odd contributions to the expectation value even if the elementary interactions of the scattering process are T-invariant. Another possibility to detect T-violation goes by comparing the transition rate of a decay and of the time reversed recombination process, obtained by exchanging the initial and final states. This is obviously an irrealistic undertaking since the entropy is strongly increased by the decay, rendering the preparation of the initial state for the recombination process too difficult. The proposal to use the entanglement between the neutral $B$ meson pairs, generated by the $\Upsilon(4S)$ decay, to tag the initial and the final states of the $B^0-\bar B^0$ oscillation \cite{banuls,wolfenstein} led finally to a convincing demonstration of the T-violating nature of the weak interactions \cite{babar}. It is believed that the T-violation of the weak interactions can not be responsible of the other time arrows. Firstly, these last-named are traced back to the initial conditions, an element of the dynamics which is independent of the equation of motion. Secondly, the current understanding of the baryogenesis requires a non-equilibrium state of the Universe, a time arrow, independently of the details of the baryon number production mechanism \cite{sakharov}. 

The role of initial conditions is obvious in breaking the time reversal symmetry even if the equation of motion is symmetric. But we may impose both the initial and final conditions in a  dynamics, described by the help of second order time derivatives, rendering the whole scheme formally more symmetric. Even if the dynamics is based on first time derivatives the second boundary condition can be imposed on the probabilities in a classical statistical system or in the quantum case, e.g. collision processes. Such a construction was proposed to achieve a manifest time reversal symmetric formalism of quantum mechanics \cite{aharonov,griffith,hartleg} and to further generalize the CTP formalism \cite{ed}. While such a scheme is well justified in the description of a small system the complexity of the final environment state of a dissipative system renders the use of final conditions unpractical.

The spontaneously broken time reversal symmetry is about irreversibility. An irreversible process is called hard or soft when it takes place at finite or vanishing frequencies, respectively. Irreversibility manifests itself perturbatively or non-perturbatively, depending on the amplitude of the motion which generates it. The non-perturbative, hard processes correspond to a large amplitude motion in finite time, such as the instability at coexisting phases or the loss of information during the collapse of the wave function and represent a specially difficult problem. We shall rather be contend to follow the building up of the soft, perturbative irreversibility, dissipation. These phenomenas are related to our limited control of large many-body systems since the observations, performed in a finite time can not fully identify the frequencies and the amplitudes. This is the origin of a non-uniform convergence and the ensuing thermodynamical time arrow is best understood by coarse graining \cite{zwanzig}. One can actually regard the effective dynamics as a result of a coarse graining in space and/or time, and this view suggests that the origin of soft irreversibility is the coarse graining of the environment. An order parameter of the time reversal symmetry is constructed below by the help of the extended action principle. The time arrow, defined by the sign of the order parameter, is  a mechanical one since it is defined without referring to a thermal bath or an equilibrium states. It corresponds to an ordered initial conditions hence its direction should agree with the thermodynamical time arrow.

The presentation starts in Section \ref{ceffs} with a brief recapitulation of the problems one faces when searching for an efficient scheme, such as the variational method, to describe classical effective theories. The spontaneous breakdown of the time reversal symmetry is described in Section \ref{spsbs} within the framework of the CTP scheme for classical, closed systems. The extension to the effective dynamics of open systems is presented in Section \ref{dynbtrs}. A brief discussion of the emergence of the finite life-time and the decoherence in open quantum systems is given in Section \ref{qchds}. The conclusions are summarized in Section \ref{concls}. There are two Appendices, added for completeness, different manifestations of the non-uniform convergence, the hallmark of spontaneous symmetry breaking, are surveyed in Appendix \ref{nonunifap}, followed by the derivation of the CTP Green function for a classical harmonic oscillator in Appendix \ref{grhos}. Readers who like to be provided with a conceptual survey before being confronted with technical details are recommended to jump now to Appendix \ref{nonunifap}.

\section{Classical effective theories}\label{ceffs}
A generic problem in classical mechanics is to find the dynamics of a coordinate $x$, called system, interacting with its environment, described by the coordinates $y=(y_1,\ldots,y_N)$ which are not followed \cite{effth,galleyeff}. We assume that the full system, described by the coordinates $x,y$, follows a time reversible and conservative dynamics which keeps the coordinates bounded at finite energy. This latter condition may require a fixed, large but finite value of the UV regulator in quantum field theoretical models. The corresponding action is written in the form $S[x,y]=S_s[x]+S_e[x,y]$, whose equations of motion are supposed to be second order differential equations in time. We need auxiliary conditions to make the solution of the equations of motion unique. They are usually initial conditions, specified for the system and the environment at the initial time, $t=t_i$, and the system trajectory is followed until a final time, $t=t_f$. The effective equation of motion, a closed equation to be satisfied by the system trajectory, is obtained in two steps. First one solves the environment equation of motion for a general system trajectory, $x(t)$. After that the solution, $y[t;x]$, is substituted into the system equation of motion. 

The effective equation of motion can be obtained as a variational equation of the effective action, $S_{eff}[x]=S[x,y[x]]$. In fact, the variational equation,
\be\label{effeom}
\fd{S_{eff}[x]}{x(t)}=\fd{S[x,y[x]]}{x(t)}+\int_{t_t}^{t_f}dt'\fd{S[x,y[x]]}{y(t')}\fd{y[t';x]}{x(t)}=0,
\ee
and the environment equation of motion,
\be\label{yeom}
\fd{S[x,y]}{y}_{|y=y[x]}=0,
\ee
imply
\be
\fd{S[x,y]}{x}_{|y=y[x]}=0.
\ee

The effective action, $S[x,y[x]]$, is not useful in irreversible systems because the local forces of the variational scheme are holonomic,
\be\label{hol}
F(x,\dot x)=-\partial_xU(x,\dot x)-\frac{d}{dt}\partial_{\dot x}U(x,\dot x),
\ee
therefore local, non-conservative dissipative forces have no place within the traditional action formalism. Furthermore, while the variational equations are derived for fixed initial and final coordinates one should rather avoid the use of final conditions in the case of a dissipative environment. Hence we need a generalization of the action principle which can handle initial condition problems, as opposed to boundary conditions, used in the traditional variation method. The trade of the final coordinates to the initial velocities is not a trivial change because the variational equation for the final coordinate, $\partial S/\partial y_f=p_f=0$, suppresses the motion. Both the local non-conservative forces and the initial conditions can be accommodated in the variational scheme by reduplicating of the degrees of freedom.

\section{A single degree of freedom}\label{spsbs}

A generalization of the action principle described in this Section handles the initial conditions as part of the dynamics. We start with the brief introduction of the generalized action principle \cite{arrow}, followed by the demonstration of the spontaneously broken time reversal invariance for the harmonic oscillator.

\subsection{Classical chronon-dynamics}\label{cctps}

We address now the two of the issues, raised in Section \ref{ceffs}, namely the need of the handling of non-holonomic forces and the use of initial conditions in the variation method of classical mechanics. The system coordinate plays a double role in the function $U(x,\dot x)$ of the holonomic forces: $\partial_xU(x,\dot x)$ and $\partial_{\dot x}U(x,\dot x)$ produce the force while $U(x,\dot x)$ represents a contribution to the energy. These two roles can be separated by the help of an active and a passive system coordinate, $x$ and $x^p$, respectively and defining a semi-holonomic force by the equation
\be\label{semihol}
F(x,\dot x)=-\partial_xU(x,\dot x,x^p,\dot x^p)_{|x^p=x}-\frac{d}{dt}\partial_{\dot x}U(x,\dot x,x^p,\dot x^p)_{|x^p=x}.
\ee
They cover the effective forces since the effective equation of motion, \eq{effeom}, can be written as
\be
\fd{S[x,y[x^p]]}{x}_{|x^p=x}=0.
\ee

\begin{figure}
\centerline{\includegraphics[width=20pc]{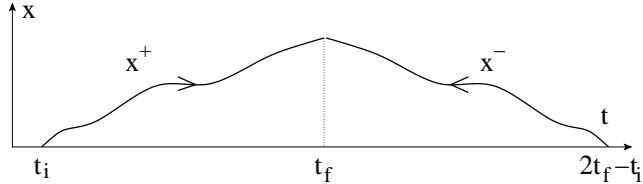}}
\caption{The system undergoes a time reversal transformation, the time arrow is flipped at $t=t_f$, and the motion is followed back to its initial state.}\label{ctppath}
\end{figure}

A new problem, left behind by the introduction of a passive copy of the system, is the treatment of this copy within the variational procedure. It can be solved, together with the problem of the initial conditions, by extending the motion for twice as long time. First we let the system follow its time evolution from the initial time $t_i$ until the final time $t_f$. After that we perform a time inversion on the state of motion and follow the reversed motion until the original initial conditions are recovered, as shown in Fig. \ref{ctppath}. The trajectory, spanned in such a manner,
\be
\tilde x(t)=\begin{cases}x(t)&t_i<t<t_f,\cr x(2t_f-t)&t_f<t<2t_f-t_i,\end{cases}
\ee
is a closed loop. We assume now that our system is subject of holonomic forces only and is governed by the Lagrangian $L(x,\dot x)$; the effective theories will be considered in Section \ref{effacts}. The action of the extended trajectory, $\tilde x(t)$, is
\be\label{naivea}
S[\tilde x]=\int_{t_i}^{t_f}dtL(\tilde x(t),\dot{\tilde x}(t))+\int^{2t_f-t_i}_{t_f}dtL^T(\tilde x(t),\dot{\tilde x}(t)),
\ee
where $T$ denotes time reversal, $\tilde x^T(t)=\tilde x(2t_f-t_i-t)$. We split the trajectory $\tilde x(t)$ into two segments, 
\be\label{doublers}
\hx(t)=(x^+(t),x^-(t))=(\tilde x(t),\tilde x(2t_f-t)),
\ee
and write
\be\label{naiveati}
S[\hx]=\int_{t_i}^{t_f}dt[L(x^+(t),\dot x^+(t))-L(x^-(t),\dot x^-(t))],
\ee
where both $x^+(t)$ and $x^-(t)$ satisfy the same initial conditions. The time reversal, $x^T(t)=x(t_f+t_i-t)$, can be represented by exchanging the trajectories, $\hx^T=\tau\hx$, where
\be
\tau=\begin{pmatrix}0&1\cr1&0\end{pmatrix}.
\ee
The CTP doublet, $\hx=(x^+,x^-)$, will be called chronon since it allows us to represent the directions of the flow of the time within the action principle and the time reversal transformation  will be called chronon conjugation. A physical trajectory, satisfying the equation of motion is  chronon conjugation invariant,
\be\label{cltraj}
x^+(t)=x^-(t).
\ee

The action \eq{naivea} is degenerate within the space of chronon conjugation invariant trajectories and in the null-space of the equation of motion of a harmonic system. We need a non-degenerate action functional when the initial conditions are handled by the action hence we add an infinitesimal piece to the CTP Lagrangian,
\be\label{clctplagr}
S[\hx]=\int_{t_i}^{t_f}dt[L(x^+,\dot x^+)-L(x^-,\dot x^-)]+S_\epsilon[\hx],
\ee
with
\be\label{splittingl}
S_\epsilon[\hx]=i\frac\epsilon2\int_{t_i}^{t_f}dt(x^{+2}+x^{-2}).
\ee
We shall see that the splitting term, $S_\epsilon$, forms an analytic structure in the frequency space to handle the non-uniform convergence arising in the long time limit around the null-space. The immediate advantage of removing the degeneracy with an imaginary term is that time reversal, the chronon conjugation, can be represented by flipping the sign of the real part of the action,
\be\label{ctpsym}
S^T[\hx]=S[\tau\hx]=-S^*[\hx].
\ee
An equivalent way to realize time reversal is to perform the transformation $t\to2t_f-t_i-t$ and $\epsilon\to-\epsilon$.

The action \eq{clctplagr} contains the doubler trajectories, $x^+(t)$ and $x^-(t)$, independently from each other and adds nothing new compared with the traditional action formalism. However the reduplication of the degrees of freedom can be used to handle time reversal non-invariant effects by the appropriate choice of the functional space of the chronon trajectory, $\hx(t)$, in which the variation is performed to derive the equation of motion: The same initial conditions, $x^\pm(t_i)=x_i$, $\dot x(t_i)=v_i$, are imposed on both doublers which are coupled at the final time by the condition
\be\label{ctpconstr}
x^+(t_f)=x^-(t_f)
\ee
which cancel the boundary contribution emerging in the calculation of the variational equation of motion. Note that $t_f$ is not a relevant parameter as long as the classical trajectory, $x(t)=x^\pm(t)$, is concerned because it is independent of the choice of $t_f$ for
$t<t_f$.

\subsection{Harmonic systems}\label{harmss}

The simplest dynamical system is defined by a quadratic action,
\be\label{hoactcn}
S[\hx]=\hf\int_{t_i}^{t_f}dtdt'\hx(t)\hat K(t,t')\hx(t')+\int_{t_i}^{t_f}dt\hx(t)\hj(t),
\ee
where $\hx=(x^+,x^-)$ and $\hj=(j_+,j_-)$ are CTP doublets and the ``metric tensor``,
\be\label{metrtens}
\hat g=\begin{pmatrix}1&0\cr0&-1\end{pmatrix},
\ee
can be used to raise or lower the chronon indices, e.g. $x_\pm=\pm x^\pm$, $j^\pm=\pm j_\pm$, in such a manner that the scalar product, $x^\sigma j_\sigma$, preserves the chronon conjugation symmetry  \eq{ctpsym} (because $\hat g\tau\hat g=-1$). The chronon Green function, $\hD=\hat K^{-1}$,  provides the classical trajectory $\hx=-\hD\hj$. If the source is physically realizable, $j_+=-j_-=j$, then both doublers follow the same classical trajectory, \eq{cltraj}, and one is left with the condition, 
\be\label{ctprelgfnct}
D^{++}+D^{--}=D^{+-}+D^{-+}.
\ee
This relation, together with the chronon conjugation symmetry and conditions $\Im j=\Im x=0$ yield the block structure
\be\label{blockg}
\Delta^{\sigma\sigma'}=\begin{pmatrix}\Delta^n&-\Delta^f\cr\Delta^f&-\Delta^n\end{pmatrix}+i\Delta^i\begin{pmatrix}1&1\cr1&1\end{pmatrix},
\ee
for both $\Delta=D$ and $\Delta=K$. The blocks $D^n$ and $D^f$ can be associated with the the near and far Green functions of the electromagnetic field. The retarded and advanced components are defined by $\Delta^{\stackrel{r}{a}}=\Delta^n\pm\Delta^f$, and the relations 
\be\label{invgrfnct}
K^{\stackrel{r}{a}}=(D^{\stackrel{r}{a}})^{-1},~~~~~~
K^i=-(D^a)^{-1}D^i(D^r)^{-1},
\ee
follow from the straightforward inversion for commutative blocks. The quadratic form of the action is symmetric, $\Delta^{\sigma\sigma'}(t,t')=\Delta^{\sigma'\sigma}(t',t)$, implying $\Delta^n(t,t')=\Delta^n(t',t)$, $\Delta^i(t,t')=\Delta^i(t',t)$, $\Delta^f(t,t')=-\Delta^f(t',t)$, and that the diagonal and the off-diagonal blocks of the harmonic action have $+1$ and $-1$ time parity, respectively. The time translation symmetry is recovered in the limit $t_f-t_i\to\infty$ where the Fourier transformation,
\be
\hat\Delta(t,t')=\int\frac{d\omega}{2\pi}e^{-i\omega(t-t')}\hat\Delta(\omega),
\ee
produces real $\Delta^n(\omega),\Delta^i(\omega)$ and imaginary $\Delta^f(\omega)$. Note that the imaginary part of the action, $K^i$, is infinitesimal and the finite imaginary part of the Green function, $D^i$, drops out in classical mechanics, playing a role in the quantum case only. 

It is instructive to check two features of the CTP formalism. The solution of the equation of motion, obtained by the help of the Green function, can be used to show the way the time inversion non-invariant forces appear in the extended action formalism. Furthermore, the argument provides the justification for $D^r$ being the retarded Green function. To start we use the parametrization $x^\pm=x\pm x^d/2$ and $j_\pm=j^d/2\pm j$ and bring the classical trajectory into the form $x=-D^rj-iD^ij^d$, $x^d=-D^aj^d$. Next, we consider the response of the extended trajectory, $\tilde x(t)$ of Fig. \ref{ctppath}, induced by a physical source, $j(t)=j_0\delta(t-t_0)$, $j^d=0$. Such a source introduces a change, $\Delta x(t)$, of the trajectory for $t_0<t<t_f$ which will be replayed in the time reversed part of the motion, $t_f<t<2t_f-t_0$. The source, $-j_0\delta(t-t_0)$, acting at that time restores the already known, initial $t_i<t<t_0$ segment of the trajectory, in particular the initial conditions, too. The lesson is that $j$ generates retarded response and preserves the chronon conjugation invariance of the trajectory. This is in agreement with the manner the variational equation selects the classical trajectory, namely both doublers start with the same initial conditions and develop the same response for $t_0<t<t_f$. Note that the final condition, \eq{ctpconstr}, is trivially satisfied and plays no particular role. This is not the case for $j^d\ne0$ which induces a more complicated response. The imaginary part plays a role only in the quantum case, and the real part, $x^d\ne0$, shows that the trajectory is non-invariant under chronon conjugation. The final condition, \eq{ctpconstr}, on the chronon trajectories plays an important role in this case. The oscillations, propagating along the doubler trajectories are reflected, turned back in time and passed to the other doubler, at the final time in such a manner that the response to $j^d$ be advanced, ie. restricted to $t_i<t<t_0$.

Another interesting point is the identification of the way the extended action formalism accommodates Newton's friction force in a harmonic oscillator. This is not possible within the framework of the traditional action principle because the time inversion parity of the antisymmetric time derivative, $\partial_t$, is $-1$ thus it generates only a boundary term in the action. However  the time derivative, $\hat\partial_t=\tau\partial_t$, is a symmetric operator due to the identity $\hat g\tau\hat g=-1$, and the action,
\be
S[\hx]=-\hf\int_{t_i}^{t_f}dt\hx(t)(m\hat\partial_t^2+m\Omega^2-k\hat\partial_t)\hx(t)+S_\epsilon[\hx],
\ee
yields the equation of motion \cite{bateman} $m\ddot x^\pm=-m\Omega^2x^\pm-k\dot x^\mp$, up to the infinitesimal terms of $S_\epsilon$. This is similar to the way the $\ord{\dddot x}$ Lorentz force emerges in electrodynamics where the role of the chronon conjugation is played by the space inversion.

\subsection{Generalized $\epsilon$-prescription}\label{ctpactreg}

The generalization of the standard $\epsilon$-prescription is presented here to handle the general auxiliary conditions, first for a harmonic oscillator, defined by the chronon action,
\be\label{lholagrc}
S[\hx;\hj]=\sum_{\sigma=\pm1}\int_{t_i}^{t_f}\left[\sigma\left(\frac{m}2\dot x^{\sigma2}-\frac{m\Omega^2}2x^{\sigma2}+j_\sigma x^\sigma\right)+i\frac{m\epsilon}2x^{\sigma2}\right].
\ee
The case of weakly interacting models will follow after the discussion of the harmonic case.

Our strategy for a harmonic oscillator is based on the use of the Green function and consists of the following two steps: First we introduce new infinitesimal quadratic terms in the action in such a manner that the solution of the equation of motion, $x=-D^rj$, automatically satisfies the trivial initial conditions, $x(t_i)=\dot x(t_i)=0$. In the next step we introduce further linear terms in the Lagrangian which generate the desired non-trivial initial conditions. 

The details of the calculation of the Green function, corresponding to the trivial initial conditions are given in Appendix \ref{grhos}, in the presence of an UV and an IR regulator, a finite time step, $\dt$, and the duration of observation, $t_f-t_i$, respectively. The initial coordinate is set to zero by the choice of the functional space, and the initial velocity is found to be vanishing for finite values of the cutoffs. The common coordinate, $z=x^+(t_f)=x^-(t_f)$, is eliminated first at a finite UV cutoff, and this step is followed by performing the continuum limit, $\dt\to0$. We find at this stage a Fourier series which represents the Green function and is non-diagonal in the frequency space, owing to the broken translation symmetry in time. The Fourier sum can be approximated by Fourier integrals for large enough $t_f-t_i$, and the Green function becomes diagonal in the frequency space, $\hD(\omega)=\hG(\omega,\Omega)/m$, as $t_i\to-\infty$, $t_f\to\infty$,
\be\label{fgrfncthof}
\hG^{\sigma\sigma'}(\omega,\Omega)=\begin{pmatrix}\frac1{\omega^2-\omega_0^2+i\epsilon}&-i2\pi\Theta(-\omega_0)\delta(\omega^2-\omega_0^2)\cr-i2\pi\Theta(\omega_0)\delta(\omega^2-\omega_0^2)&-\frac1{\omega^2-\omega_0^2-i\epsilon}\end{pmatrix},
\ee
or equivalently,
\be\label{grefreq}
G^n=P\frac1{\omega^2-\Omega^2},~~~
G^f=-i\pi\sign(\omega)\delta(\omega^2-\Omega^2),~~~
G^i=-\pi\delta(\omega^2-\Omega^2),
\ee
as long as the inequalities
\be\label{acdbs}
t_f-t\ll\frac\Omega\epsilon\ll t_f-t_i,
\ee
are assured. The first inequality protects the final condition against the eroding effects of the  $\epsilon$-prescription, the second bound assures that the discrete sum of the Fourier series can be approximated by the Fourier integral and places us deeply in the non-uniform limit, displayed in Fig \ref{arrf}. 

The Green function has off-shell and on-shell components, the latter denoting the modes of the null-space of the kernel of the harmonic action. The symmetric part, $D^n$, given by the principal value prescription contains the response of the harmonic oscillator to the external source and represents the off-shell part. The rest, $D^f$, and $D^i$, are on-shell, the former takes care of the initial conditions and the latter is responsible of the interference effects in the quantum case, cf. Section \ref{qchds}.

After the Green function, $\hD$, generating the solution with trivial initial conditions has been found we turn our attention to the case of non-trivial initial conditions. Since the limits  $t_i\to-\infty$ and $t_f\to\infty$ are performed in the construction of the Green function we replace the auxiliary conditions, imposed on some fixed time, by the requirement that the trajectory be of the form
\be\label{homsol}
x_a(t)=x_0\sin(\Omega t+\alpha)
\ee
for sufficiently large $-t$ or $t$, respectively. The generalized $\epsilon$-prescription is defined by the action 
\be\label{geneps}
S[\hx]=\hf\int_{-\infty}^\infty dtdt'[\hx(t)-\hx_a(t)]\hat K(t-t')[\hx(t')-\hx_a(t')],
\ee
where $\hat K=\hD^{-1}$, $\hD$ being given by eq. \eq{fgrfncthof} with $\epsilon=0^+$ and $\epsilon=0^-$ for initial and final conditions, respectively, and
with $x^\sigma_a=x_a$. It is easy to see that the equation of motion produces the desired trajectory and the inversion of the Green function yields $\hat K(\omega)=m\hat H(\omega,\Omega)$, where
\be\label{hogrinv}
H^{\sigma\sigma'}(\omega,\Omega)=(\omega^2-\Omega^2)\begin{pmatrix}1&0\cr0&1\end{pmatrix}+i\epsilon\begin{pmatrix}1&-2\Theta(-\omega)\cr2\Theta(\omega)&-1\end{pmatrix},
\ee
in particular
\be\label{invprop}
H^n=\omega^2-\Omega^2,~~~H^f=i\sign(\omega)\epsilon,~~~H^i=\epsilon.
\ee
The action \eq{lholagrc} can be written in the form \eq{clctplagr} in the limit $t_i=-\infty$, $t_f=\infty$ with
\be\label{splittingli}
S_\epsilon[\hx]=\epsilon\left[\frac{i}2\int_{-\infty}^\infty dt[x^{+}(t)-x^{-}(t)]^2+\frac1\pi P\int_{-\infty}^\infty dtdt'\frac{x^+(t)x^-(t')}{t-t'}\right],
\ee
where $P$ denotes the principal value prescription. 

A few remarks are in order at this point:
\begin{enumerate}
\item The coupling at the final time within a chronon, \eq{ctpconstr}, is transformed into an infinitesimal, time translation invariant coupling. 

\item The initial conditions are represented by the homogeneous solution of the equation of motion which belong to the null-space of the equation of motion operator. They influence the action of the generalized $\epsilon$-prescription in $\ord{\epsilon}$ which is enough to generate a finite effect via the null-space singularity of \eq{modedec}.

\item Since the initial conditions are handled by the imaginary terms of the action the time flows in such a the direction which makes $\Im S[\hx]$ increasing during the motion, cf. \eq{ctpsym}.
\end{enumerate}

\begin{figure}
\centerline{\includegraphics[width=20pc]{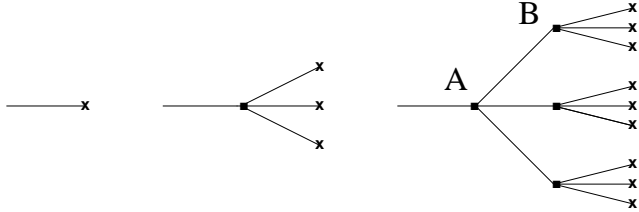}}
\caption{The iteration of the equation of motion \eq{intmeom} can be visualized by tree-graphs. Here the lines represent the Green functions, $D^r$, the dots stand for the vertex $g$ and the crosses represent the source, $j$.}\label{treegr}
\end{figure}

The aforementioned analysis can be extended to weakly interactive systems. Let us consider the model described by the coordinate $x=(x_1,\ldots,x_N)$ and the action
\bea\label{intact}
S[\hx]&=&\sum_n\int_{-\infty}^\infty dtdt'[\hx_n(t)-\hx_{an}(t)]m_n\hat K(t-t',\Omega_n)[\hx_n(t')-\hx_{an}(t')]\nn
&&-\frac14\sum_\sigma\sigma\sum_{n_1n_2n_3n_4}g_{n_1n_2n_3n_4}\int_{t_i}^{t_f}dtx^\sigma_{n_1}(t)x^\sigma_{n_2}(t)x^\sigma_{n_3}(t)x^\sigma_{n_4}(t).
\eea
The solution of the equation of motion,
\be\label{intmeom}
[\hD^{-1}_n(\hx_n-\hx_{an})]^\sigma=\sum_{n_2n_3n_4=0}^Ng_{nn_2n_3n_4}x_{n_2}^\sigma x_{n_3}^\sigma x_{n_4}^\sigma-j^\sigma_n,
\ee
with $\hD^{-1}_n(\omega)=m_n\hat H_n(\omega,\Omega_n)$, can be found by iteration, the tree-graphs of the first three orders are depicted in Fig. \ref{treegr}. The homogeneous solutions can be taken into account by the shift of the external source, $\hj_n\to\hj_n+(\hD_n)^{-1}\hx_{hn}$. One can prove by the repeated application of the equation $\sum_{\sigma'}\sigma'D^{\sigma\sigma'}=D^r$ that the CTP Green functions, $\hD_n$, can be replaced by $D_n^r$ in the equation of motion. The result is the equation of motion
\be
(D^r_n)^{-1}(x_n-x_{hn})=\sum_{n_2n_3n_4=0}^Ng_{nn_2n_3n_4}x_{n_2}x_{n_3}x_{n_4}-j_n
\ee
for physical trajectory with the external source, $j^\sigma_n=j_n$. 

To avoid the secular terms we apply the adiabatic switching approximation, the interactions are turned on gradually in time, $g\to g(t)$, where the coupling strength is negligible at the initial time, $g(-\infty)=0$. The time dependence in the coupling constant spreads the discrete spectrum contributions, generated by the iteration at integer multiples of $\Omega$, into the continuous spectrum.

\subsection{Mechanical time arrow}

The auxiliary conditions break the time reversal symmetry independently of the equation of motion, the dynamics. However the breaking of the symmetry seems to be part of the dynamics when the auxiliary conditions are represented in the chronon action by $L_\epsilon$. The regulated retarded Green function, \eq{regretgf}, converges in a non-uniform manner during the removal of the null-space regulator, $T\to\infty$, c.f. Fig. \ref{arrf}, and amplifies the infinitesimal symmetry breaking, \eq{invprop}, to finite effects, \eq{grefreq}. This is reminiscent of the non-uniform convergence of the order parameter in the thermodynamical limit when spontaneous symmetry breaking takes place except that now the symmetry breaking is in time rather than in space. This similarity can further be corroborated by constructing an order parameter for the time reversal symmetry.

A localized, time reversal invariant external perturbation, $j(t)=j_0\delta(t-t_0)$, generates a response, $\delta x(t)$, which is not time reversal invariant,  $\delta\dot x(t_0)\ne0$, due to the boundary conditions in time. Since the transformation $\epsilon\to-\epsilon$ exchanges the initial and final conditions the discontinuity at $\epsilon=0$, 
\be\label{locopar}
\chi(t_0)=-\frac1{2j_0}\lim_{\dt\to0}\lim_{\epsilon\to0^+}[\dot x_\epsilon(t_0+\dt)+\dot x_\epsilon(t_0-\dt)]-(\epsilon\to-\epsilon),
\ee
serves as a local order parameter of time reversal symmetry where the $\epsilon$-dependence is displayed explicitly. This transforms into a time independent order parameter,
\be\label{globordptr}
\chi=-2\dot D^f(0),
\ee
in the limits $t_i\to-\infty$ and $t_f\to\infty$ and acquires the value $\chi=1/m$ for the Green function \eq{dnfi}. 

The construction, presented for the trivial initial condition, remains well defined for arbitrary initial conditions. In fact, the $\epsilon$-dependence appears through the quadratic part of the action and the initial conditions, encoded by the linearly coupled source, do not modify the order parameter. Thus 
\be
\tau_{mech}=\sign(\chi)=\sign(\epsilon)
\ee
can be interpreted as a mechanical time arrow, defined for a single degree of freedom without referring to environment or equilibrium state.

As to the regulator dependence of the order parameter the retarded Green function \eq{regretgf} yields $\chi=F(\epsilon T)$ where the function $F(z)$ is defined by eq. \eq{ordparh}. The order parameter shows the hallmarks of spontaneous symmetry breaking, namely the non-commutativity of the removal of two effects, the IR cutoff and the external symmetry breaking. When this is carried out in the proper order, the symmetry breaking is removed in an infinitely large system, a finite symmetry breaking is left over.

\section{Open systems in the thermodynamical limit}\label{dynbtrs}

Let us suppose that a variation in time of the system trajectory, $\delta x(t)$, generates a perturbation, $\delta y(t)$, of the environment trajectory. The non-vanishing order parameter, \eq{locopar}, considered for the environment indicates that $\delta y(t)$ is not time reversal invariant therefore its reaction back to the system, $\delta x'(t)$, breaks the time reversal symmetry, as well, in such a manner that the order parameter is determined by the environment time arrow only. Such a transmutation of the environment time arrow to the system, to be demonstrated in details in this Section, originates from the $\epsilon$-prescription of the environment and can be considered as a spontaneous breakdown of the time reversal symmetry. The difference between the realization of the symmetry breaking in a closed and an open systems is that the action has $\ord\epsilon$ and $\ord{\epsilon^0}$ symmetry breaking terms in the former and the latter case, respectively. In the case of infinitely large environment with continuous spectrum soft perturbative  irreversibility follows for certain system initial conditions. If the environment has gapless modes then an all initial conditions lead to irreversible dynamics.

To separate the symmetry breaking within the system and in the environment it is advantageous to allow to set independently the initial or the final conditions for the system and the environment. This freedom allows us to introduce two independent $\epsilon$ parameters, $\epsilon_s$ and $\epsilon_e$ for the system and the environment action, respectively which induce two order parameters by applying the definition \eq{locopar} separately to the system and the environment.

\subsection{Effective chronon theory}\label{effacts}

The application of the CTP scheme to a closed, conservative system amounts to a simple paraphrase of the standard equations and the power of this formalism manifests itself within the effective theories of open systems \cite{effth}. The chronon action,
\be\label{clctpact}
S[\hx,\hy]=S[x^+,y^+]-S[x^-,y^-]+S_\epsilon[\hx]+S_\epsilon[\hy],
\ee
of the full, closed system yields the effective action,
\be\label{effacte}
S_{eff}[\hx]=S[x^+,y[\hx]]-S[x^-,y[\tau\hx]]+S_\epsilon[\hx].
\ee
The trajectory $y[\hx]$ is the solution of the environment equation of motion,
\be
\fd{S[\hx,\hy]}{\hy}=0,
\ee
and satisfies the environment auxiliary conditions which are initial conditions for dissipative environments. Note that owing to the independent variations of $y^+(t)$ and $y^-(t)$, the solution, $y[x^+,x^-]$, contains more information than $y[x]$, arising from solving eq. \eq{yeom}. According to Fig. \ref{ctppath} the couplings between the members of a chronon, $\delta^2 y[\hx]/\delta x^+\delta x^-$, arise from the coupling of the trajectories at the final time, $t_f$. The couplings which survive the limit $t_f\to\infty$ correspond to phenomena which decouple from the system and take place in the asymptotic long time state of the environment. 

One can separate the original system dynamics in the effective action by writing $S_{eff}[\hx]=S_s[\hx]+S_{infl}[\hx]$ where 
\be
S_{infl}[\hx]=S_e[x^+,y[\hx]]-S_e[x^-,y[\tau\hx]]
\ee
denotes the influence functional \cite{feynman}. Another way to write the effective action is
\be\label{ekchrc}
S_{eff}[\hx]=S_1[x^+]-S_1[x^-]+S_2[\hx],
\ee
where $\delta^2S_2[\hx]/\delta x^+\delta x^-\ne0$. The interpretation of $S_1$ and $S_2$ can be inferred from the equations of motion. It is specially advantageous to write these equations by using the parametrization $x^\pm=x\pm x^d/2$ \cite{keldysh} because it is sufficient to calculate the effective action in $\ord{x^d}$ for chronon conjugation invariant trajectories satisfying eq. \eq{cltraj}. The variational equation for $x^d$ at $x^d=0$,
\be
0=\fd{S_1[x]}{x}+\fd{S_2[x,x^d]}{x}_{|x^d=0},
\ee
indicates that that the one-point action, $S_1$, includes the conservative part of the effective dynamics, and the two-point action, $S_2$, covers non-conservative, semiholonomic forces. 

We can find the same effective action by Legendre transformation, as well, by borrowing the ideas from quantum mechanics and defining the generator functional,
\be\label{legtr}
W[\hj]=S[\hx,\hy]+\int_{t_i}^{t_f}dt\hj(t)\hx(t),
\ee
where the system and the environment coordinates are eliminated by their equation of motion,
\be\label{fseom}
\fd{S[\hx]}{\hx}+\hj=\fd{S[\hx]}{\hy}=0.
\ee
The $n$-point Green functions are given by the functional Taylor expansion,
\be
W[\hj]=\sum_{n=0}^\infty\frac1{n!}\int_{t_i}^{t_f}dt_1\cdots dt_nD^{\sigma_1,\ldots,\sigma_n}(t_1,\ldots,t_n)j_{\sigma_1}(t_1)\cdots j_{\sigma_n}(t_n),
\ee
and can be found by iteration. The system trajectory is actually the one-point function,
\be\label{trajfder} 
x^\sigma(t)=\fd{W[\hj]}{j_\sigma(t)}_{|\hj=0},
\ee
and the higher order Green functions describe the dependence of the trajectory on the source, given in terms of the higher order graphs in the iterative solution of the equation of motion, c.f. Fig. \ref{treegr}. The effective action, \eq{effacte}, is recovered as the Legendre transform of $W[\hj]$,
\be\label{invlegtr}
S_{eff}[\hx]=W[\hj]-\int_{t_i}^{t_f}dt\hj(t)\hx(t),
\ee
where $\hj$ is eliminated on the right hand side by inverting eq. \eq{trajfder}.

\subsection{Energy balance}\label{enballs}

The recovery of the symmetry with respect to translation in time as $t_f-t_i\to\infty$ has been proven in Section \ref{ctpactreg}. This symmetry is a characteristic feature of soft irreversibility where the energy loss to the environment in equilibrium is due to the uncontrollable small amplitude, slow fluctuations rather than external time dependence in the dynamics. We turn our attention to the energy balance equation to discover the dissipative effective forces.

We start with the harmonic oscillator, \eq{harmosclagr}, coupled to an external source, $L\to L+jx$. The work done by the oscillator on the external source,
\be
W=-\int dtj(t)\dot x(t),
\ee
can be expressed as a frequency integral,
\be
W^{\stackrel{r}{a}}=i\int\frac{d\omega}{2\pi}\omega j(-\omega)[D^{\stackrel{r}{a}}(\omega)j(\omega)+x_a(\omega)],
\ee
where the upper and the lower case refers to the initial and the final conditions, respectively. The coefficient of the Green function in the integrand is an odd function of the frequency and suppresses the the contribution of $D^n$, allowing the replacement $D^{\stackrel{r}{a}}\to\pm D^f$, $D^f$ being given by eq. \eq{grefreq}. Since the support of $D^f$ is the null-space the energy exchange in an asymptotically long time takes place in the null-space only,
\be
W^{\stackrel{r}{a}}=\pm\frac{|j(\Omega)|^2}{2m}+\frac{x_0\Omega}2\Im[j(\Omega)e^{-i\alpha}].
\ee
This is a natural result since only the null-space modes can be non-vanishing in the absence of the external source and can accumulate the energy excess or loss. Note that $\sign(W)=\sign(\chi)$ for $x_0=0$, in agreement with the possibility of using the direction of time in which the energy of the oscillator is lost as an order parameter of the time reversal invariance in dissipative dynamics.

The conserved quantities, defined formally in the CTP formalism by the help of the Noether theorem, are vanishing due to the chronon conjugation symmetry \eq{ctpsym}. Hence we have to perform the symmetry transformations on one copy only, leaving the other, representing the environment, unchanged in deriving the CTP analogy of Noether's theorem \cite{effth,galleyeff,fhydro}. The conserved quantities are defined by the help of $S_1[x]$, and if $S_2[\hx]$ violates a symmetry then the corresponding conservation law receives a non-vanishing source term and becomes a balance equation. 

Let us now assume that the effective action corresponds to an effective Lagrangian, $L_{eff}(\hx)=L_1(x^+)-L_1(x^-)+L_2(\hx)$, containing time derivatives up to order $n_d$. The equation of motion for $x^+$ is
\be
\sum_{j=0}^{n_d}(-1)^j\frac{d^j}{dt^j}\fd{}{x^{(j)}}[L_1(x^+)+L_2(x^+,x^-)]_{|x^-=x^+}=0,
\ee
where the notation $x^{(j)}=d^jx/dt^n$ is used. To find the energy balance equation we make a  variation, $\delta x^+=-\xi\dot x$, $\delta x^-=0$ and write the linearized action of $\xi(t)$,
\be
S[\xi]=-\sum_{j=0}^{n_d}\int dt\frac{d^j\xi\dot x}{dt^{(j)}}\fd{L}{x^{(j)}},
\ee
in the form
\bea
S[\xi]&=&\int dt\biggl\{\xi\left[\partial^-_tL_2(\hx)-\frac{d}{dt}L_1(x^+)\right]\nn
&&-\sum_{j=0}^{n_d}\left(\sum_{k=1}^j(^j_k)\fd{L_1(x^+)}{x^{+(j)}}\xi^{(k)}x^{+(j-k+1)}+\sum_{k=0}^j(^j_k)\fd{L_2(\hx)}{x^{+(j)}}\xi^{(k)}x^{+(j-k+1)}\right)\biggr\}_{|x=x^+=x^-},
\eea
after some partial integrations, where the partial derivative $\partial^-_t$ acts on $x^-$ only. The corresponding equation of motion for $\xi$, when $x=x^+=x^-$ is the classical trajectory, is
\be\label{enballe}
\dot H_{n^d}=\kappa,
\ee
where
\be
H_1=\fd{L_1}{\dot x}\dot x-L_1
\ee
for $n_d=1$ and
\be\label{effen}
H_n=H_1+\sum_{j=2}^{n_d}\left[j\fd{L_1}{x^{(j)}}x^{(j)}+\sum_{k=1}^{j-1}(^j_{k+1})(-1)^k\frac{d^k}{dt^k}\left(\fd{L_1}{x^{(j)}}x^{(j-k)}\right)\right]
\ee
for $n_d\ge2$ with the source term,
\be
\kappa=-\sum_{j=0}^{n_d}\fd{L_2}{x^{-(j)}}_{|x^+=x_-=x}x^{(j+1)},
\ee
representing that part of the interaction energy which is lost to the environment at the final time. The rate of change is not definite for large amplitude and fast motion but the bound $\kappa\le0$ will be proven in the next Section for perturbative soft irreversibility.

There is a fundamental difference between the energy of the Newtonian mechanics and $H_n$, namely the latter is not definite for $n_d\ge2$ \cite{ostrogadsky}. The effective dynamics is usually non-local in time if the frequency spectrum is discrete and the local gradient expansion assumes continuous spectrum, ie. an infinite environment. The conservation of the total, system plus the environment energy can not prevent that the system energy becomes unbounded from below in this case.

\subsection{Normal modes I.: Finite system}\label{spectrcs}

We now extend the argument of the previous section to a multi-dimensional harmonic system, described by the coordinate $z=(x,y_1,\ldots,z_N)$ and the action,
\be\label{normdact}
S=\hf\sum_{n,n'}\int_{-\infty}^\infty dtdt'\hat z_n(t)\hat K_{nn'}(t-t')\hat z_{n'}(t')+\int_{-\infty}^\infty dt\hat z_1(t)\hj(t).
\ee
describing time reversal invariant and conservative forces, and seek the effective action for $\hx=\hat z_1$ which can be calculated by the help of the generator functional \eq{legtr}. All normal modes have the same mechanical time arrow in the model. First we diagonalize the action by the orthogonal transformation,
\be
z_n=\sum_jA_{nj}w_j.
\ee
transforming the normal modes, $w$, into $z$. Next, the generator functional,
\be\label{wharsys}
W[\hj]=\sum_j\left[\hf\int_{-\infty}^\infty dtdt'\hat w_j(t)\hat k^{-1}_j(t-t')\hat w_j(t')+A_{1,j}\int_{-\infty}^\infty dt\hat w_j(t)\hj(t)\right],
\ee
with
\be
\hat k_j=\sum_{n,n'}A_{nj}\hat K_{nn'}A_{n'j},
\ee
is constructed by solving the normal mode equations of motion and inserting the solution into $W[\hj]$,
\be
W[\hj]=-\hf\sum_j\int_{-\infty}^\infty dtdt'\hj(t)A_{1j}\hat k^{-1}_j(t-t')A_{1j}\hj(t').
\ee
Finally, one arrives at the effective action,
\be\label{nmefact}
S_{eff}[\hx]=\hf\int_{-\infty}^\infty dtdt'\hx(t)\hat K(t-t')\hx(t')+\int_{-\infty}^\infty dt\hx(t)\hj(t),
\ee
with
\be\label{effhatnmet}
\hat K^{-1}=\sum_jA^2_{1j}\hat k^{-1}_j.
\ee

The solution of the effective equation of motion, $\hx=-\hD j$, is obtained in terms of the Green function, $\hD=\hat K^{-1}$. We use a physically realizable source, $j_\sigma=\sigma j$, giving rise to the response $x^\pm=x=-D^rj$ where $D^r=(K^r)^{-1}$, according to the inversion rules \eq{invgrfnct}. The equation of motion satisfied by this trajectory is $K^rx=-j$ with
\be
\frac1{K^r(\omega)}=\sum_j\frac{A^2_{1j}}{M_j[\omega^2-\Omega_j^2+i\sign(\omega)\epsilon]},
\ee
where $M_j$ and $\Omega_j\ge0$ denote the $j$-th normal mass and frequency, respectively and $\sum_jA^2_{1j}=1$. 

The structure of the effective equation of motion can better be understood by noting that the external source, $j$, is coupled to several normal modes, cf. eq. \eq{wharsys} hence the energy injected into the system by the source is spread over the normal modes and generates a rather complicated response, reminiscent of interacting systems \cite{hydro}. The retarded Green function,
\be\label{reteomor}
D^r(\omega)=\frac1{K^r(\omega)}=\sum_j\frac{A^2_{1j}}{M_j}\left[\frac{\omega^2-\Omega_j^2}{(\omega^2-\Omega_j^2)^2+4\epsilon^2}-i\sign(\omega)\frac{2\epsilon}{(\omega^2-\Omega_j^2)^2+4\epsilon^2}\right],
\ee
given for $\epsilon\ne0$, shows that $\Re K^r(\omega)$ is non-vanishing and time reversal invariant except in the normal mode spectrum. At the same time $\Im K^r(\omega)$ is localized in the frequency space within the $\epsilon$ neighborhood of the spectrum and breaks the time reversal invariance and $\sign(\omega\Im K^r(\omega))=\sign(\omega\epsilon)$. The discrete peaks of $\Im K^r(\omega)$ represent the breakdown of the time reversal invariance in the effective dynamics, induced by the environment auxiliary conditions.

\subsection{Normal modes II.: Infinite system}\label{contspects}

It is advantageous to introduce the spectral function,
\be
\rho(\Omega)=\sum_j\frac{A^2_{1j}}{2M_j\Omega_j}\delta(\Omega_j-\Omega),
\ee
for infinite systems, $N=\infty$, yielding 
\bea\label{reteom}
\frac1{K^r(\omega)}&=&\sum_j\int d\Omega\delta(\Omega_j-\Omega)\frac{A^2_{1j}}{M_j}\frac1{(\omega+i\epsilon)^2-\Omega^2}\nn
&=&\int d\Omega\frac{2\rho(\Omega)\Omega}{(\omega+i\epsilon)^2-\Omega^2},
\eea
where the possible dependence on the order of performing the summation and the integration is revealed by the non-uniform convergence in the second line as $\epsilon\to0$. 

As a simple example take a mixed spectrum,
\be\label{spectrd}
\rho(\Omega)=\frac1{2m_s\omega_s}\delta(\Omega-\omega_s)+\rho_D(\Omega),
\ee
where the environment is represented by a Drude-type spectral weight, 
\be\label{drude}
\rho_D(\Omega)=\Theta(\Omega)\frac{g^2\Omega}{m\Omega_D(\Omega_D^2+\Omega^2)}.
\ee
The roots of
\be
K^r(\omega)=\frac{m_s[(\omega+i\epsilon)^2-\omega_s^2]}{1-\frac{\pi g^2m_s}{\Omega_Dm}\frac{(\omega+i\epsilon)^2-\omega_s^2}{\Omega_D-i\omega}},
\ee
$\omega^{(\pm)}=\pm\omega_s-i\epsilon$ and $\omega^{(0)}=-i\Omega_D$, define the spectrum of an irreversible effective theory. 

To understand better the dependence of the first equation in \eq{reteom} on the order of the summation and the integration, let us make a coarse graining in time by using the trajectories $x(t)\to x_T(t)=x(t)c_T(t)$, where $c(t)\approx1$ if $|x|\ll T$ and $c(t)\approx0$ if $|t|\gg T$, $T$ being an IR cutoff function with time reversal invariance, $c(-t)=c(t)$. Such a restriction of the trajectories amounts to the spread
\be
x_T(\omega)=\int\frac{d\omega'}{2\pi}x(\omega-\omega')c_T(\omega')
\ee
of the discrete frequency lines in the frequency space. We introduce the minimal separation of the discrete normal frequencies, $\Delta\tilde\omega=\inf|\tilde\omega_j-\tilde\omega_{j'}|$, and distinguish the following cases \cite{dynbr}:
\begin{itemize}
\item $\Delta\tilde\omega>0$: The observations, carried out in time $T\gg1/\Delta\tilde\omega$, can resolve all normal modes and the effective theory for $x_T$ is conservative. We can reproduce such observations with an effective theory where the integration over the spectral variable is performed first, followed by the summation. The linear equation of motion operator is given by \eq{reteomor} and the motion is reversible at frequencies which do not belong to the normal mode spectrum.
 
\item $\Delta\tilde\omega=0$: The normal frequency spectrum has an accumulation point and the long but finite time measurements leave infinitely many normal modes unresolved. The infinitely many unresolved normal modes act as an uncontrollable absorber of the system energy and the effective dynamics for $x_T$ with finite $T$ contains dissipative forces. The summation must be carried out first in the first equation in \eq{reteom} in this case, leaving the integration for the second step.
\end{itemize}

\begin{figure}
\centerline{\includegraphics[width=20pc]{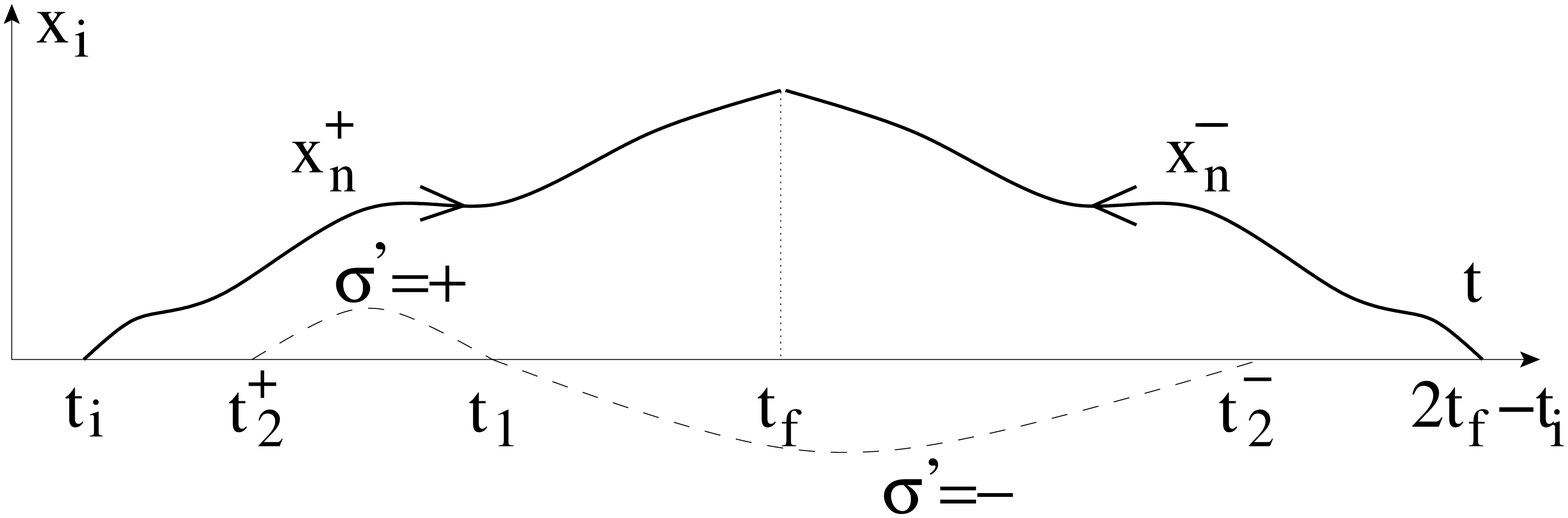}}
\caption{The system coordinate, $x_1$, acts as an external source on the environment coordinate $x_n$, $n>1$, c.f. Fig. \ref{ctppath}. The dashed lines represent $D_{n\sigma'}^+(t_1,t_2)z_1^{\sigma'}(t_2)$, $t_2^-=t_2$, $t_2^-=2t_f-t_i-t_2^+$, the response of $z_n(t_1)$ on $z_1(t_2)$.}\label{ctppathint}
\end{figure}

The soft irreversibility can be found in the iterative solution of weakly interactive models, too. Let us take for instance the model described by the action \eq{intact}, whose iterative solution can be written as a series of tree-graphs, as shown on Fig. \ref{treegr}. Consider the contribution of the Green function, $D^r_n=D^n_n+D^f_n$, $n>1$, corresponding to the line AB in the third graph. The system coordinate acts as an external source on the environment coordinate $x_n$ and the contribution $D_{n\sigma'}(t_1,t_2)x_1^{\sigma'}(t_2)$, to $x^\sigma_n(t_1)$, shown by dashed lines in Fig. \ref{ctppathint}, represents the oscillations of $x_n$ at $t_f=\infty$, the absorber for the system energy. 

The leading order solution with trivial environment initial conditions is
\be
x_n^\sigma(t)=g_{n000}\sum_{\sigma'}\int_{-\infty}^\infty dt'D^\sigma_{~\sigma'n}(t-t')x^{\sigma'3}_1(t')+\ord{g^2},
\ee
for $n>1$ and its insertion into the chronon action yields the influence functional for $x=x_1$,
\be
S_{infl}[\hx]=\hf\sum_{\sigma\sigma'}\sigma\sum_{n=1}^Ng^2_{n000}\int_{-\infty}^\infty dtdt'x^{\sigma3}(t)D_{\sigma\sigma'n}(t-t')x^{\sigma'3}(t')+\ord{g^3}.
\ee
The spectral function,
\be
\rho(\Omega)=\sum_n\frac{g^2_{n000}}{2m_n\omega_n}\delta(\Omega-\omega_n),
\ee
allows us to write
\be
S_{infl}[\hx]=\sum_{\sigma\sigma'}\int d\Omega\Omega\rho(\Omega)\int_{-\infty}^\infty dtdt'x^{\sigma3}(t)G_{\sigma\sigma'}(t-t',\Omega)x^{\sigma'3}(t')+\ord{g^3},
\ee
where the Fourier transform of $\hG(t-t',\Omega)$ is given by eq. \eq{fgrfncthof}. The corresponding effective equation of motion for $x^+=x^-$ is
\be
\ddot x(t)=-\omega_0^2x(t)+x^2(t)\frac6{m_0}\int d\Omega\Omega\rho(\Omega)\int_{-\infty}^\infty dt'G^r(t-t',\Omega)x^3(t')+\ord{g^3}.
\ee
Although there are nonlinear terms, the discussion of the breakdown of the time reversal invariance and the emergence of dissipative forces, presented for the harmonic model, applies.

\subsection{Toy model}\label{toymods}

We can better isolate the role of the environment in generating irreversibility by looking into the harmonic model of the Lagrangian \cite{caldeira},
\be\label{holagr}
L=\frac{m_0}2\dot x^2-\frac{m_0\omega^2_0}2x^2+\sum_{n=1}^N\left(\frac{m_n}2\dot y_n^2-\frac{m_n\omega^2_n}2y_n^2-g_ny_nx\right)
\ee
where the time may run in different direction in the system and the environment, for $x$ and $y$, respectively. The chronon action assumes the form
\be
S[\hx,\hy]=\hf\int_{-\infty}^\infty dtdt'[\hx(t)\hD^{-1}_0(t-t')\hx(t')+\hy(t)\hG^{-1}(t-t')\hy(t')]-\int_{-\infty}^\infty dt\hx(t)g\hy(t),
\ee
in terms of the trajectories $\hx=\tau^{\Theta(-\epsilon_s)}\hx$, $\hy=\tau^{\Theta(-\epsilon_e)}\hy$ and $g=\sign(\epsilon_s\epsilon_e)\mr{diag}(g_1,\ldots,g_n)$. 

The elimination of the environment leads to the effective action, 
\be
S[\hx,\hy]=\hf\int_{-\infty}^\infty dtdt'\hx(t)\hat K(t-t')\hx(t'),
\ee
where the inverse Green function, $\hat K=\hD_0^{-1}-\hat\Sigma=\hD^{-1}$, contains the self energy $\hat\Sigma=g\hat g\hG\hat g g$, where $\hat g$ is given by \eq{metrtens}. The perturbative inversion yields the geometric series,
\be\label{schwdtoy}
\hD=\hD_0+\hD_0\hat\Sigma\hD_0+\hD_0\hat\Sigma\hD_0\hat\Sigma\hD_0+\hD_0\hat\Sigma\hD_0\hat\Sigma\hD_0\hat\Sigma\hD_0+\cdots.
\ee
When applied to the source, $j_\sigma=\sigma j$, both $\hD_0$ and $\hat\sigma$ can be replaced by their retarded component,
\be\label{schwdret}
D^r=D^r_0+D^r_0\Sigma^rD^r_0+D^r_0\Sigma^rD_0^r\Sigma^rD_0^r+D^r_0\Sigma^rD_0^r\Sigma^rD_0^r\Sigma^rD_0^r+\cdots.
\ee

The influence functional,
\be\label{cleinflf}
S_{infl}[\hx]=-\hf\int_{-\infty}^\infty dtdt'\int\frac{d\omega}{2\pi}e^{-i\omega(t'-t)}\hx(t')\hat\Sigma(\omega)\hx(t),
\ee
can conveniently be parametrized by means of the spectral function,
\be\label{spectrtoy}
\rho(\Omega)=\sum_n\frac{g_n^2}{2m_n\omega_n}\delta(\omega_n-\Omega),
\ee
and the self energy assumes the form
\be
\hat\Sigma(\omega)=\int_0^\infty d\Omega2\Omega\rho(\Omega)\hG(\omega,\Omega),
\ee
where $\hG(\omega,\Omega)$ is given by eq. \eq{fgrfncthof}, in particular
\bea\label{selfetoy}
\Sigma^n(\omega)&=&2P\int_0^\infty d\Omega\frac{\Omega\rho(\Omega)}{\omega^2-\Omega^2},\nn
\Sigma^f(\omega)&=&-\sign(\epsilon_e)i\pi\sign(\omega)\rho(|\omega|),\nn
\Sigma^i(\omega)&=&-\sign(\epsilon_e)\pi\rho(|\omega|).
\eea
The expansion of the exponent in the right hand side of eq. \eq{cleinflf} in $t'-t$ and $\hat\Sigma(\omega)$ in $\omega$ yields the influence Lagrangian,
\be\label{toyinflf}
L_{infl}=-\hf(x\vec\Sigma^nx^d+x^d\vec\Sigma^nx+x^d\vec\Sigma^fx-x\vec\Sigma^fx^d+x^di\vec\Sigma^ix^d),
\ee
where
\be
\vec{\hat\Sigma}=\sum_{\ell=0}^\infty\frac{(-1)^\ell}{\ell!}\partial^\ell_{i\omega}\hat\Sigma(0)\partial_t^\ell.
\ee

The equation of motion, generated by the variation of $x^d$, at $x^d=0$, is
\be\label{toyem}
m\ddot x=-(m\omega^2_0+\vec\Sigma^r)x,
\ee
with $\vec\Sigma^r=\vec\Sigma^n+\vec\Sigma^f$ and contains the following terms up to $\ord{\partial_t^2}$: (i) $\ord{\partial_t^0}$: a normal frequency renormalization, $\omega^2_0\to\omega^2_0-\Delta\omega^2$, with
\be
\Delta\omega^2=\frac2m\int_0^\infty d\Omega\frac{\rho(\Omega)}\Omega,
\ee
(ii) $\ord{\partial_t}$: a Newtonian friction force, $F=-k\dot x$, with $k=\pi\rho'(0)$, and (iii) $\ord{\partial_t^2}$: a mass renormalization, $m\to m+\delta m$, where 
\be
\delta m=4\int_0^\infty d\Omega\frac{\rho(\Omega)}{\Omega^3}.
\ee

The finite imaginary part of the self energy dominates the imaginary part of the effective action making $\chi_s=0$ and $\sign(\chi_e)=\sign(\epsilon_e)$. The first equation indicates that no symmetry breaking is taking place within the system and the second equation shows that the mechanical time arrow of the environment is simply transfered to the system since $\tau_{mech~s}=-\sign\dot D^f(0)$. This is reminiscent of spontaneous symmetry breaking, namely the replacement $\epsilon\to h$, and $T\to V$ brings the order parameter \eq{globordptr} into $M$, mentioned in Section \ref{spsymbrs}. 

One can separate the following cases:
\begin{enumerate}
\item Discrete environment spectrum without condensation point: The breakdown of the time reversal symmetry is not universal. The experiments, performed in a sufficiently long time reveal the time reversal invariant system dynamics, realized at frequencies which do not belong to the environment spectrum. 

\item Discrete environment spectrum with condensation point: The continuous spectral function becomes a good approximation for finite time observations at the condensation point and indicate the irresistible loss of energy to those modes. 

\item The continuous environment spectrum: We find soft irreversibility at any frequency. The analysis of the discrete spectrum reveals that the loss of energy is due to the environment, whose modes are degenerate with the system.
\end{enumerate}

The soft irreversibility is the result of the mixing of the environment modes into the system dynamics, described by $\Sigma^f$. This mixing can clearly be seen in the geometric series \eq{schwdret}, where $\Sigma^r=\Sigma^n+\Sigma^f$ and each $\Sigma^f$ stands for environment excitations. These excitations decouple from the system after the system-environment interactions is adiabatically switched off because they correspond to the environment null-space.

It is instructive to calculate the energy balance by using the results of Section \ref{enballs} for continuous spectrum. Let us assume the validity of the expansion of the effective action in the time derivative and use the effective Lagrangian,
\be
L_{eff}=\hf(x^+K^nx^+-x^-K^nx^-+x^+K^fx^--x^-K^fx^+),
\ee
where $K^n=\sum_jK^n_j\partial_t^{2j}$ and $K^f=\sum_jK^f_j\partial_t^{2j+1}$. After some partial integration we have
\be
L_{eff}=\sum_j(-1)^j\left\{\hf K^n_j[(x^{+(j)})^2-(x^{-(j)})^2-K^f_jx^{-(j)}x^{+(j+1)}\right\},
\ee
and the corresponding energy balance equation, \eq{enballe}, contains the system energy
\be\label{energy}
H=-\hf K^n_0x^2-\hf K_1^n\dot x^2+\sum_{j=2}^n(-1)^jK^n_j\left\{\left(j-\hf\right)x^{(j)2}+\sum_{k=1}^{j-1}(^j_{k+1})(-1)^k\frac{d^k}{dt^k}[x^{(j)}x^{(j-k)}]\right\},
\ee
and the source term
\be
\kappa=\sum_{j=0}^n(-1)^jK^f_jx^{(j+1)2}.
\ee
The total energy loss, the integral of $\kappa$ during the motion, allows us to define the energy loss,
\be\label{enloss}
\kappa(\omega)=\sum_{j=0}^n(-1)^jK^f_j(\omega)\omega^{2(j+1)}|x(\omega)|^2,
\ee
for each frequency.

Consider first the Green function in the form of the normal mode decomposition, c.f. \eq{reteomor}, where $D^f(\omega)$ is imaginary and the positivity of the norm in the coordinate space, $z$, guarantees the inequalities
\be\label{spectrdef}
\sign(\omega)iD^f(\omega)\ge0,
\ee
and $\sign(\omega)\Im K^f(\omega)>0$. The latter yields $(-1)^j\sign(\omega)K^f_j(\omega)\le0$ and makes the system energy non-increasing, $\dot H\le0$, within each frequency mode. Naturally this result holds only for the trivial environment initial conditions and it is easy to see that the source term, $\kappa$, is non-definite for non-trivial initial conditions. Let us now return to the toy model where the time arrow can formally be set separately for the system and the environment. Since $\sign(\omega)\sign(\Im(K^f(\omega))=\sign(\epsilon_e)$ according to the second equation of \eq{selfetoy} we have $\sign(\omega)\sign(iD^f(\omega))=\sign(\epsilon_e)$, indicating that the system dynamics is stable  and the system energy is non-increasing whatever direction the time flows for the system if that direction is the same as for the environment time.

\section{Finite life-time and decoherence}\label{qchds}

We turn to the quantum case where the breakdown of time reversal invariance induces genuine quantum effects. The soft irreversibility stands for the possibility of any mode of the system to loose energy to the environment, leading to the leakage of the norm of the system state into the environment, the dynamical origin of the finiteness of the life-time of the system excitations, in the quantum case. Another effect of the gapless excitation spectrum is that the slightest system-environment interaction generates a system-environment mixing according to the degenerate perturbation expansion, and the system state becomes mixed by decoherence. This represents an irreversible loss of the informations, contained in the suppressed off-diagonal elements of the density matrix.

\subsection{Quantum chronon-dynamics}

The CTP formalism was introduced by J. Schwinger for the perturbation expansion of the expectation values, 
\be\label{expvqm}
\la\psi(t_i)|U^\dagger(t,t_i)AU(t,t_i)|\psi(t_i)\ra,
\ee
in quantum mechanics \cite{schw}. Here $|\psi(t_i)\ra$ denotes the state at the time $t_i$ and $U(t,t')$ is the time evolution operator. The reduplication of the degree of freedom originates from the simultaneous presence of the bra and ket in the expectation value, two states, developing in opposite direction in time but representing the same physical system.

The naive quantization of the chronon dynamics of a closed system, introduced in Section \ref{cctps}, is based on the Hamiltonian, $H=\hat p\dot{\hx}-L$, the replacement of c-numbers with operators, $[x^\sigma,p_{\sigma'}]=\delta^\sigma_{\sigma'}i\hbar$, and the representation of the chronon location $\hx=(x^+,x^-)$ by the help of the dyadic product, $|x^+\ra\la x^-|$. The use of the dyadic products rather than the kets eliminates a disadvantageous feature of the bracket formalism, namely the dependence of the action of anti-linear operators on the choice of the basis. An anti-linear operator, $A$, acts as a complex conjugation on c-numbers, $Ac|\psi\ra=c^*A|\psi\ra$, hence its action on a basis vector depends whether this latter is represented by real of complex functions, an inherent ambiguity in a linear space over the complex numbers. In fact, by assuming $|\psi\ra=c|\chi\ra$ we have 
\be
Aa|\psi\ra=\begin{cases}a^*c|\chi\ra&A|\psi\ra=|\psi\ra,\cr a^*c^*|\chi\ra&A|\chi\ra=|\chi\ra.\end{cases}
\ee
The usual convention in representing time inversion in quantum mechanics by an anti-linear operator, $T$, is to impose $T|x\ra=|x\ra$. By assuming $|\psi^\pm\ra=c_\pm|\chi^\pm\ra$ we find in the linear space of states of chronons,
\be
Aa|\psi^+\ra\la\psi^-|=\begin{cases}a^*c^*_+c_-|\chi^-\ra\la\chi^+|&A|\psi^+\ra\la\psi^-|=|\psi^-\ra\la\psi^+|,\cr a^*c^*_+c_-|\chi^-\ra\la\chi^+|&A|\chi^+\ra\la\chi^-|=|\chi^-\ra\la\chi^+|.\end{cases}
\ee

The time evolution of the chronon wave function, $\psi(\hx)$, can be obtained by the help of a path integral expression,
\be\label{chrpint}
A(t_f,\hx_f,t_i,\hx_i)=\int D[\hx]e^{\ih S[\hx]},
\ee
where the integration is over trajectories with end points $\hx(t_i)=\hx_i$, $\hx(t_f)=\hx_f$ and the action in the exponent is given by eqs. \eq{clctplagr}-\eq{splittingl}. The standard procedure to derive the equation of motion for $A$, which usually gives the Schr\"odinger equation, now leads to the Neumann equation, indicating that $\psi(t,\hx)=A(t,\hx,t_i,\hx_i)$ is actually the density matrix. This is not surprising after noting the formal similarity between the doubling $x\to(x^+,x^-)$ and $\psi(x)=\la x|\psi\ra\to\rho(x^+,x^-)=\la x^+|\rho|x^-\ra$ in classical and quantum systems, respectively. Such an interpretation reveals a fundamental difference between $\rho(\hx)$ and $\psi(x)$: The expectation value, $\la A\ra_\rho$, is additive in the quantum state, $\la A\ra_{\rho_1+\rho_2}=\la A\ra_{\rho_1}+\la A\ra_{\rho_2}$, without interference. The state $\rho$ contains all quantum effects, i.e. all interference has already been placed into $\rho(\hx)$. The only expression for $\la A\ra_\rho$ which is additive in $\psi$ and respects the condition \eq{ctpconstr} of virtual variation is $\la A\ra_\rho=\Tr[\rho A]$. The time reversal, realized either by chronon conjugation or the transformation $\epsilon\to-\epsilon$ in classical physics, is represented by the complex conjugation in the coordinate basis.

From now on we return to the traditional quantization scheme and introduce the quantum version of the generator functional \eq{legtr}, 
\be\label{ctpgf}
e^{\ih W[\hj]}=\Tr T[e^{-\ih\int dt(H(t)-j^+(t)x(t))}]\rho_iT^*[e^{\ih\int dt(H(t)-j^-(t)x(t))}],
\ee
where the trace stands for the the condition \eq{ctpconstr} and $\rho_i$ denotes the initial density matrix with the path integral representation
\be\label{pintcgf}
e^{\ih W[\hj]}=\int D[\hat x]e^{\ih S[\hat x]+\ih\int dt\hx(t)\hj(t)}.
\ee

\subsection{Open quantum systems}\label{oqss}

If the observed system is not closed but interacts with its environment, the action $S[x,y]=S_s[x]+S_e[x,y]$, leads to the path integral expression,
\be\label{ctpeffgf}
e^{\ih W[\hj]}=\int D[\hat x]e^{\ih S_{eff}[\hat x]+\ih\int dt\hx(t)\hj(t)},
\ee
where the bare (Wilsonian) effective action, $S_{eff}[\hx]=S_s[x^+]-S_s[x^-]+S_{infl}[\hx]$, where the influence functional is given by
\be\label{qinfl}
e^{\ih S_{infl}[\hx]}=\int D[\hy]e^{\ih S_e[x^+,y^+]-\ih S_e[x^-,y^-]+\ih S_\epsilon[\hy]},
\ee
and the decomposition \eq{ekchrc} can be used again.

It is instructive to consider a generalization of the CTP formalism, based on the density matrix,
\be
\rho(x^+_f,x^-_f)=\la x^+_f|U(t_f,t_i)\rho_iU^\dagger(t_f,t_i)|x^-_f\ra,
\ee
rather than its trace, as in eq. \eq{ctpgf}. It is the Open Time Path scheme, and the path integral expression for the reduced density matrix is
\be
\rho(x^+_f,x^-_f)=\int D[\hx]D[\hy]e^{\ih S[x^+,y^+]-\ih S[x^-,y^-]+\ih S_\epsilon[\hx]+\ih S_\epsilon[\hy]},
\ee
with $\hx(t_f)=\hx_f$ and $y^+(t_f)=y^-(t_f)$, and the effective action of the CTP formalism gives
\be\label{otpedm}
\rho(x^+_f,x^-_f)=\int D[\hx]e^{\ih S_1[x^+]-\ih S^*_1[x^-]+\ih S_2[\hx]+\ih S_\epsilon[\hx]}.
\ee
The decoherence in the coordinate diagonal representation, the suppression of the contributions of well separated chronon trajectories, is driven by $\Im S_2$. The couplings between the doubler trajectories in $S_2$ which generate semiholonomic forces in classical mechanics now stand for the contributions of several final environment states in the trace of \eq{ctpgf}, make the system state mixed and represent the system-environment entanglement. In particular, the $\ord{\hx^2}$ terms in \eq{splittingli} represent an infinitesimal decoherence and the corresponding initial state is mainly the ground state but contains an infinitesimal amount of mixing.

The two trajectories of a chronon are identical in classical physics, c.f. eq. \eq{cltraj}. It is easy to see that the unitarity of the time evolution preserves eq. \eq{cltraj} on the level of the averages. In fact, the average of the coordinate at time $t_o<t_f$ can be calculated in two equivalent, $t_f$-independent manners, 
\be
\Tr[x\rho(t_o)]=\fd{W[\hj]}{j^\pm(t)},
\ee
owing to the identities
\bea
\la\psi(t_i)|U^\dagger(t_o,t_i)xU(t_o,t_i)|\psi(t_i)\ra&=&\la\psi(t_i)|U^\dagger(t_f,t_i)U(t_f,t_o)xU(t_o,t_i)|\psi(t_i)\ra\nn
&=&\la\psi(t_i)|U^\dagger(t_o,t_i)xU^\dagger(t_f,t_o)U(t_f,t_i)|\psi(t_i)\ra,
\eea
c.f. the remark about the $t_f$ independence of the classical trajectory, made at the end of Section \ref{cctps}. This allows us to identify $x^d=x^+-x^-$ with the quantum fluctuations in the coordinate basis.

The reduplication of the degrees of freedom deals with time reversal symmetry breaking interactions within the framework of the action formalism. It seems to be only a formal device in classical mechanics because the equation of motion make the doublers within a chronon equivalent. This changes when quantum systems are considered: Eq. \eq{cltraj} holds in the average only and the dynamical independence of the doublers within a chronon are needed to accommodate the principle of superposition. In fact, the uncertainty principle excludes the use of the classical trajectory to extract the dynamics of the momentum which is rather found by the help of the higher moments of $x^d$, the violation of eq. \eq{cltraj}, in the path integral \eq{pintcgf}. In other words, the quantum fluctuations are the microscopic manifestation of the presence of the dynamically independent doublers. The time arrow is opposite for the the bra and the ket, the two members of the chronon and generate advanced effects of the quantum fluctuations. In fact, consider two degrees of freedom, described by the coordinates $x_1$ and $x_2$, coupled linearly by the term $gx_1x_2$ of Lagrangian. The product $gx_2$ acts as a linear source for $x_1$ and the remarks, made in Section \ref{harmss}, show that the effect of $j^d=j_++j_-\to gx_2^d$ is advanced. The opposite orientation of the time arrow for bras and ket makes the time arrow for $x=(x^++x^-)/2$ and $x^d=x^+-x^-$ opposite, as well, and has an important role, it produces positive Lyapunov exponent for the quantum fluctuations.

\subsection{Propagator}

The generator functional, \eq{ctpgf}, motivates the introduction of the generalized time ordered product \cite{umezawa}, 
\be
\bar T[A^\sigma(t)B^{\sigma'}(t')]=\begin{cases}\Theta(t-t')A^\sigma(t)B^{\sigma'}(t')+\Theta(t'-t)B^{\sigma'}(t')A^\sigma(t)&(\sigma,\sigma')=(+,+),\cr
\Theta(t'-t)A^\sigma(t)B^{\sigma'}(t')+\Theta(t-t')B^{\sigma'}(t')A^\sigma(t)&(\sigma,\sigma')=(-,-),\cr
A^\sigma(t)B^{\sigma'}(t')&(\sigma,\sigma')=(-,+),\cr
B^{\sigma'}(t')A^\sigma(t)&(\sigma,\sigma')=(+,-),
\end{cases}
\ee
and the definition of the propagator
\be
\Tr\bar T[x^\sigma(t)x^{\sigma'}(t')]=i\hbar D^{\sigma\sigma'}(t,t'),
\ee
which turns out to be identical with the classical Green function, \eq{fgrfncthof} in the case of the harmonic oscillator and ground state initial conditions. The diagonal block gives the Feynman propagator, $D_F=D^{++}$, and the off diagonal part defines the spectral function, given by the Wightmann function,
\bea\label{spectrfnt}
iD^{-+}(\omega)&=&\Tr[A(-\omega)A(\omega)]\nn
&=&\sum_{mn}\la m|A(-\omega)|n\ra\la n|A(\omega)|m\ra\nn
&=&\sum_{mn}|\la m|A(\omega)|n\ra|^2\nn
&\ge&0,
\eea
where $|n\ra$ are eigenstates of the Hamiltonian and the last inequality follows from the positivity of the norm. The spectral function is vanishing for $\omega<0$ if the motion starts in the ground state and $iD^i(\omega)=\sign(\Re\omega)D^f(\omega)$ follows, or equivalently, the Feynman propagator can be reconstructed by the help of the retarded or advanced Green functions,
\be
D_F(\omega)=D^n(\omega)+iD^i(\omega)
=\begin{cases}D^r(\omega)&\Re\omega>0,\cr D^a(\omega)&\Re\omega<0.\end{cases}
\ee
Since $D^r(\omega)$ is a real function of $i\epsilon$ an equivalent form of this equation is
\be\label{rafeynm}
D^{\stackrel{r}{a}}(\omega)=D_F(\pm|\Re\omega|+i\Im\omega).
\ee
Similar relations hold for the inverse Green function, $iK^i(\omega)=\sign(\omega)K^f(\omega)$, furthermore the inequality \eq{spectrfnt} assures the bound \eq{spectrdef}.

It is illuminating to rewrite the harmonic action \eq{hoactcn}, in terms of $x^\pm=x\pm x^d/2$ and $j^\pm=\pm j+j^a/2$,
\bea
S[x,x^d]&=&\hf\int_{t_i}^{t_f}dtdt'[x^d(t)K^r(t,t')x(t')+x(t)K^a(t,t')x^d(t')+x^d(t)iK^i(t,t')x^d(t')]\nn
&&+\int_{t_i}^{t_f}dt[x(t)j^a(t)+x^d(t)j(t)],
\eea
showing that the auxiliary variable, $j^a$, generates the $\ord{\hbar^0}$ expectation value of the coordinate when $j_+=-j_-\ne0$ in \eq{pintcgf}, expressed by the help of the real components of either the action or the propagator, $D^r=(K^r)^{-1}$ and $D^a=(K^a)^{-1}$, which satisfies the classical equation of motion. A genuine $\ord\hbar$ quantum effect, the linear superposition of quantum fluctuations, is displayed by the imaginary component, $D^i=-D^rK^iD^a$ which drops out from the equations of classical mechanics. 

The imaginary part of the propagator manifests itself both in the diagonal and the off-diagonal CTP blocks of the propagator which controls two different quantum effects. The imaginary part of the diagonal blocks of the action, $\Im K^{\pm\pm}$, modifies the normalization of a pure state or the trace of the density matrix. It is well known that the Hilbert space provides a ray representation of the physical states, namely the kets $|\psi\ra$ and $c|\psi\ra$ correspond to the same physical state. The electromagnetic interaction arises as a relativistic gauge theory, based on the redundancy of a pure phase, $|c|=1$. The redundancy of $|c|\ne1$ opens the possibility of the leakage of the state to another sector of the Hilbert space, the finite life-time of a state. The diagonal block, $K^{++}$, is the inverse of the Feynman propagator when the system starts in the ground state \cite{effth}, and its roots, $D_F^{-1}(\omega_0)=0$, correspond to the normal mode frequencies. Thus the time scale of the leakage of the system into the environment, $1/|\Im\omega_0|$, a result of the system-environment entanglement, is the damping time scale of the classical trajectory according to eq. \eq{rafeynm}. Note that the unitarity of the time evolution is preserved in the CTP effective theories which keep the trace of the reduced density matrix unity hence a quantum state with finite life-time is mixed, reflecting the system-environment entanglement. Another quantum effect, the decoherence in coordinate representation, is the result of the imaginary parts of both the diagonal and the off-diagonal blocks. In fact, they contribute to the $\ord{x^{d2}}$ term of the effective action and suppress the contribution of the chronon trajectories to the reduced density matrix in \eq{otpedm} with well separated doublers. The identity $\Im K^{++}=\Im K^{-+}$ implies that the intrinsic time scale of of the decoherence and the dissipation are identical in a harmonic system. 

Both the mixing of the state and the decoherence are already present with an infinitesimal strength even in a closed system. In fact, the first and the second term in the action, \eq{splittingli}, of the generalized $\epsilon$-prescription represent an infinitesimal decoherence and mixing, respectively. Both become of finite strength in the effective theory for an infinitely large environment with continuous spectrum via the spontaneous breakdown of the time reversal symmetry. 

The identity of the classical and quantum Green functions makes the generalization of the classical treatment of harmonic models to the quantum case trivial. The quantum version of the toy model of Section \ref{toymods} has already been throughly examined \cite{grabert,hupaz}. The calculation of the CTP effective theory proceeds as in the classical case, in particular the influence Lagrangian \eq{toyinflf} can be recovered and the classical equation of motion remains valid for the expectation value of the coordinate. The non-increasing nature of the energy, given by eq. \eq{energy}, is assured by the positivity of the norm and the Schwinger-Dyson resummation, \eq{schwdtoy}, can be interpreted as a result of the mixing of the degenerate system and environment excitations. Similar procedure yields the quadratic effective Lagrangian for a test particle, interacting with an ideal gas \cite{frict}. The quadratic classical effective Lagrangian of the point charge \cite{point} remains valid in non-relativistic QED, too.

\subsection{Double role of the environment induced time arrow}\label{taps}

While the CTP action of a closed system, \eq{clctplagr}, has an infinitesimal imaginary part the effective action, \eq{effacte}, may acquire finite imaginary pieces, cf. the third equation of \eq{selfetoy}. The initial conditions of the trajectories $x^\pm(t)$ are identical in classical physics, making $x^d=x^+-x^-=0$ and keeping the imaginary terms of the equation of motion for $x=(x^++x^-)/2$ infinitesimal. The initial conditions are more involved in the quantum case where the quantum fluctuations show up already in the initial state, $x^d(t_i)\ne0$, and the finite imaginary part of the effective action displays two genuine $\ord\hbar$ quantum effects, the finiteness of the life-time and decoherence:

\begin{itemize}
\item The transfer of the mechanical time arrow from the environment to the system by the spontaneous breakdown of the time reversal symmetry determines the direction of the time in which the dissipative work, performed by the system is positive and the motion is stable. The norm of the system state is non-increasing in the time direction.

\item The positive norm of the system states makes $\Im S_2\ge0$, where $S_2$ is defined by the help of the decomposition \eq{ekchrc} of the effective action and renders the interference among the possible system histories destructive rather than constructive in the same time direction. 
\end{itemize}

The mixing of states in the system and the environment sectors and the decoherence change the norm of the state during the time evolution. Both effect are kept stable by the transfer of the environment time arrow in a manner similar to the stabilization of the classical energy, discussed in Section \ref{toymods}.

\section{Conclusions}\label{concls}

The role of the environment initial conditions in generating soft irreversibility was studied in this work. Our particular scheme, the CTP formalism, is chosen to assure consistency since the environment initial conditions play a crucial role in the effective dynamics, and it is appropriate  to incorporate all initial conditions into the dynamics. This way of encoding the initial conditions leads to null-space divergences to be regulated by an $\epsilon$-prescription. The frequency integrals converge in a non-uniform manner during the removal of the cutoff, $\epsilon\to0^+$, a feature that generates the spontaneous breakdown of the time reversal symmetry within a gapless, harmonic environment. The spontaneous symmetry breaking, a phase transition, takes place only if the number of degrees of freedom tends to infinite. How can such a phenomenon be observed in a harmonic model where the normal modes are non-interacting hence independent? The resolution of the apparent puzzle is that if the system coordinate is coupled to infinitely many environment normal modes then a small modification of the coupling may produce a large, singular change in the system dynamics. While each environment normal mode generates infinitesimal time reversal asymmetric effective interactions they make together the effective dynamics irreversible if their spectral weight is sufficiently large.

The local, dissipative effective forces have been derived in the classical CTP formalism for harmonic and weakly coupled systems and the energy balance equation has been obtained. The energy is found to be non-increasing for each frequency mode in a harmonic system if the environment develops from a stable, stationary state. 

The approach, pursued in this work, provides an order parameter to monitor the status of the time reversal symmetry on mechanical ground. The mechanical time arrow, defined by the sign of the order parameter, agrees with the thermodynamical time arrow because the initial system, satisfying some fixed initial condition, has minimal entropy. The spontaneous symmetry breaking is usually the result of the increased sensitivity of symmetry breaking fluctuations in an unstable equilibrium point. While this picture can not completely be carried over the symmetry breaking within the time the formal similarities between spontaneous symmetry breaking and the transfer of the mechanical time arrow suggests the emergence of a unique time arrow within large, closed, stationary systems.

The present approach leaves several questions open. The different levels of the breakdown of the time reversal invariance has been established by means of the $\epsilon$-prescription and, quite obviously, they depend on the details of our approach. Is the resulting scenario generally valid, independent of the formalism or we see just some non-physical features of a particular analytic structure on the complex frequency plane? Another question concerns the relation between the thermodynamical and the quantum time arrows, between the soft and the hard irreversibility in general. Which features of soft irreversibility remain valid for irreversible processes taking place at a fixed scale, like the one, in Wilson's cloud chamber?

\acknowledgments{Acknowledgments}

It is pleasure to thank J\'anos Hajdu for enlightening discussions and the encouragement during the years the results, presented here, have been accumulated and the careful reading of the manuscript.

\appendix

\section{Non-uniform convergence and symmetry breaking}\label{nonunifap}

The large environment, defined by a cutoff, may produce counter intuitive effective dynamics because the convergence during the removal of the cutoff is non-unform. The origin and the physical implications of different kinds of non-unfiform space-time convergences are brefly reviewed in this appendix.

\subsection{Long distance cutoff: phase transition}\label{spsymbrs}

\begin{figure}
\centerline{\includegraphics[scale=1]{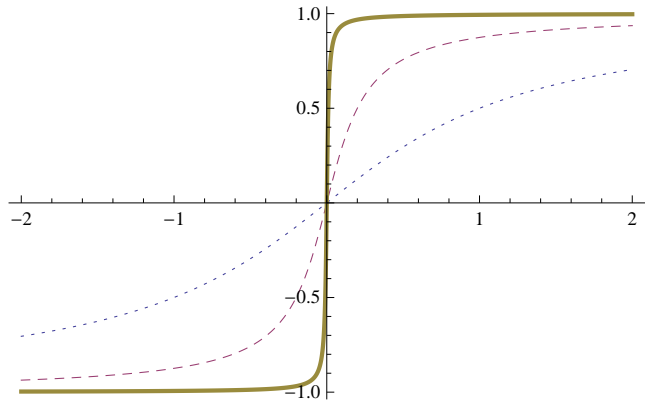}}
\caption{The qualitative dependence of the order parameter $M$ on the explicit symmetry breaking, $h$, for $V=1$ (dotted line), $V=5$ (dashed line) $V=100$ (thick line).}\label{magnf}
\end{figure}

The simplest way to control the number of degrees of freedom of a system, $N$, is the use of a large but finite size, $L$. The infinite system is defined by the thermodynamical limit, $L\to\infty$, performed by keeping thedensity $N/V$, $V=L^3$ fixed. The spontaneous symmetry breaking can be detected in a sufficiently large system by measuring averages in stationary equilibrium states. One introduces an external, explicite symmetry breaking of strength $h$ and checks the status of the symmetry in the limit $h\to0$. A qualitative volume dependence of the order parameter of the symmetry $h\to-h$ is given by $M(V,h)=F(Vh)$, with
\be\label{ordparh}
F(z)=\frac1\pi\Im\ln\frac{1+iz}{1-iz},
\ee
where the volume $V$ controls the closeness of a pole to the real, physical $h$ axis, cf. Fig \ref{magnf}. The survival of the symmetry breaking as $h\to0$ indicates the existence of a long characteristic time scale and the non-commutativity of the $h\to0$ and the thermodynamical limits. 

The dynamical basis of the non-commutativity is a diverging intrinsic time scale: Observations, carried out in arbitrary large, but finite time in an infinitely large system may give different results than the infinitely long time observations in large, but finite system. The result is the coexistence of different formal measures in the statistical ensembles of infinite systems, a non-trivial phase structure. The choice of a phase at the coexisting point by the dynamics is a new feature, realized by strictly infinite systems only.

It is easy to follow this phenomenon in a special class of phase transitions, the spontaneous symmetry breaking which is due to the slowing down of the order parameter. Let us consider a macroscopic, solid body, consisting of $N$ particles of mass $m$ which interact with translation and rotation invariant forces and separate the translational and rotational motion by distinguishing laboratory and body-fixed, co-moving coordinate systems, the latter is defined by having vanishing center of mass velocity and a diagonal tensor of inertia
\be
\Theta^{jk}=m\sum_nx^j_nx^k_n.
\ee
The collective coordinates $\v{X}$, $\theta,\phi,\alpha$ consist of the vector of the translation and the Euler angles of the rotation which bring the laboratory frame into the body-fixed coordinate system. These variables serve as order parameters for translations and rotations and their effective dynamics is described by the Hamiltonian
\be\label{hamrot}
H_{coll}=\frac1{2M}\v{P}^2+\hf L^j(\Theta^{-1})^{jk}L^k,
\ee
written in terms of the total momentum and angular momentum, $\v{P}$ and $\v{L}$, respectively. 

The different representations of a finite Heisenberg algebra are unitary equivalent and the ground state of the Hamiltonian \eq{hamrot} is a singlet with vanishing expectation value for the  order parameters. But the excitation spectrum becomes dense for large systems and the classical description of  the order parameters becomes an excellent approximation for macroscopic bodies. The result is the slowing down of the order parameters: Assuming that the energy of each degrees of freedom is $k_BT/2$ in a pure state the order of magnitude of the velocity of the collective coordinates of a system of mass $M$, given in grams, and characteristic size $L$, expressed in centimeters, is $\dot X\approx\sqrt{T/M}\times10^{-8}$ cm/s and $\dot\theta,\dot\phi,\dot\alpha\approx\sqrt{T/M}/L\times10^{-8}$/s, where the temperature is given in Kelvin. The order parameters can safely be considered as stationary on the macroscopic scale. 

The importance of the non-commutativity of the thermodynamical and the long observation time limits is not restricted to the status of symmetries and such a singularity characterizes the phase transitions without order parameter, too. A typical example is the fluid-vapor transition which can not be related to symmetry breaking since one can arrive from one phase to the other without encountering singularities by going around the tricritical point.

\subsection{Short distance cutoff: dynamically modified relations}\label{uvans}

The cutoff which controls the number of degrees of freedom, placed close to each other in space, is a minimal distance, $a$. Its removal, the renormalization procedure, may produce non-uniform convergence and thereby induce unexpected violation of some relations, believed to be safe of modification by the interaction.

The easiest way to locate the origin of the non-uniform convergence in the renormalization of quantum field theories is the BPHZ renormalization scheme \cite{bogoljubov,hepp,zimmermann} which can briefly be summarized as follows. One starts with the formal perturbation series of the one particle irreducible Green functions and modifies the Lagrangian in a recursive manner in each order of the perturbation expansion to introduce a regulated, well defined theory which converges as $a\to0$. Let us start with the contribution of a Wick-rotated Feynman graph, given in the Euclidean space-time, to a one-particle irreducible vertex function in the form of the loop-integral,
\be\label{bhhzli}
G(p)=a^{[G]}\int d\tilde qI(\tilde p,\tilde q,\tilde g),
\ee
where $p$ and $q$ denote the external momentum and the integration variables, respectively and $g$ stands for the parameters of the Lagrangian. The quantities with tilde are made dimensionless by the cutoff, $\tilde q=aq$, $\tilde g=ga^{-[g]}$, etc. The Wick rotation, the analytic extension of the loop-integral for imaginary energy, is needed to decouple the problem of null-space  singularities, considered later, from the large $q$ divergences, treated here. The loop integral \eq{bhhzli} is only formal because it may be divergent. The overall divergence of this integral is the divergence which comes from the integration domain where all components of $q$ diverge with the same rate as $a\to0$. The power counting is a simple algorithm to find the degree of the overall divergence, it is given by the mass dimension of the integral, $[G]$, $\hbar=c=1$. The loop integrals with $[G]<0$ are finite and need no special attention. The itegrals with overall divergence, $[G]\ge0$ are dangerous and one modifies the Lagrangian by adding a counterterms for each such graph. The counterterms are defined by the first $[G]$ order of the Taylor expansion of $G$ in $p$ around $p=0$ and their impact on the perturbation series is the subtraction of the Taylor expansion terms from the graph in question. As a result, the graph \eq{bhhzli} is now well defined and assumes its bare, regulated value, $G_B(p)=G(p)-P(p)$, where $P(p)$ is polynomial of $p$, consisting of the first $[G]$ order of $G(p)$. The complicated part of the BPHZ scheme is a rather involved, recursive proof that the removal of the overall divergences at the order of the perturbation expansion where they appear leaves no sub-divergences behind and makes the limit $a\to0$ convergent and finite. 

The distinguished feature of the BPHZ scheme is that it can be carried out before the integration, by directly subtracting the divergent terms from the integrand \cite{zimmermann}. The subtracted loop-integral has overall degree of divergence $-1$ therefore it is finite and plays no role anymore in the renormalization. This makes this scheme attractive because it allows us to remove the cutoff before the integration and the renormalized Feynman graphs can be given in terms of finite integrals, without the explicite use of the cutoff. Such a removal of the divergences is realized by the use of appropriately chosen, cutoff-dependent parameters to define the bare Lagrangian, $\tilde g\to\tilde g_a$, and the renormalized loop-integral is defined by 
\be\label{subtrli}
G_R(p)=\lim_{a\to0}a^{[G]}\int d\tilde qI(\tilde p,\tilde q,\tilde g_a).
\ee

We now turn to the question of uniform convergence. The uniform convergence of the loop-integral is important because it allows us to interchange the order of the removal of the cutoff, $a\to0$, and the integration and thereby to define the renormalized perturbation expansion in terms of integrals, written without any cutoff. Furthermore it makes the loop-integral independent of the order of the integration over the energy and the momentum. A simple way to show that the integral \eq{subtrli} converges uniformly is to find an integrable bound, i.e. a function $F(p,q)\ge a^{[G]}|I(\tilde q,\tilde p,\tilde g_a)|$, with
\be
\int dqF(p,q)<\infty.
\ee
It is easy to check that the Wick-rotated loop-integrals with negative overall degree of divergence converge uniformly. 

It may happen that there are graphs in the theory with non-negative overall degree of divergence which are accidentally finite \cite{bardeen}. These graphs are finite and need no counterterms but may converge in a non-uniform manner and make the renormalized theory, defined by the limit $a\to0$, different from the theory which is obtained by simply setting $a=0$ in the subtracted integrand. Such a surprising phenomenon, a cutoff leaves a finite trace in the dynamics even after it has been removed, is called in a somehow unfortunate manner anomaly. Physical phenomenas, such as the anomalous breakdown of chiral invariance and the neutral pion decay \cite{adler,bell}, the breakdown of scale invariance close to the critical point, the emergence of the proton mass in QCD \cite{marciano} and the Abraham-Lorentz force \cite{point} owe their existence to such a  restricted convergence of the renormalization procedure. 

The space-time continuum of the renormalized theories possesses features beyond the Bolzano-Weierstrass theorem. These are characterized by the (finite) counterterms of the accidentally finite vertex functions, an unexpected source of free parameters in the Lagragnian. The impact of them on the dynamics can not be localized in the space-time hovewer they change relations which are thought to be safe of effects of the interactions. It is a difficult task to find these relations, and much simpler to figure out the differences by inspecting equations expressing some symmetries. This might be the reason for non-uniform convergence during the removal of the UV cutoff is often refered to as the dynamical symmetry breaking.

Phase transitions and spontaneous symmetry breaking are usually driven by the potential energy, the kinetic energy being less important for the long distance modes, with a remarkable exception, the deconfinement transition in finite temperature Yang-Mills theories \cite{deconf}. The dynamical symmetry breaking and the singularities of the one-particle sector, reviewed below, are also related to the kinetic energy.

\subsection{Short time cutoff: propagation along fractals}

The description of a single degree of freedom needs infinitely many modes in time. The discretization of the time with the time step $\dt$ maximizes the frequency of observing the system at $\Lambda=2\pi/\dt$ and allows us to approach the continuous time in a mathematically well defined manner. It will be shown below that such an approach leads to nowhere differentiable, fractal trajectories for a non-relativistic point particle, a feature needed to respect the canonical commutation relation.

We shall work with a harmonic oscillator, defined by the Lagrangian,
\be\label{harmosclagr}
L=\frac{m}2\dot x^2-\frac{m\Omega^2}2x^2.
\ee
and consider the trace of the time evolution operator,
\be
f(t_f-t_i,\dt)=\Tr_\dt\left[e^{-\ih H(t_f-t_i)}\right],
\ee
regulated for long time by a large, but finite propagation time, $t_f-t_i$, and for short time by $\dt$. The path integral expression,
\be
f(t_f-t_i,\dt)=\left(\frac{m}{2\pi i\hbar\dt}\right)^{\frac{N}2}\prod_{j=1}^N\int dx_je^{\ih\dt\sum_{\ell=1}^N[\frac{m}2(\frac{x_\ell-x_{\ell-1}}2)^2-\frac{m\Omega^2}2x_\ell^2]},
\ee
where $x_0=x_N$ and $\dt=(t_f-t_i)/N$ gives rise to the propagator
\be\label{propsum}
\Tr_\dt[T[x(m\dt)x(n\dt)]]=\frac{i\hbar}m\frac{\Delta\omega}{2\pi}\sum_{k=1}^N\frac{e^{-i\frac{2\pi}Nk(m-n)}}{\frac4{\dt^2}\sin^2\frac{\omega_k\dt}2-\Omega^2+i\frac\epsilon{m\dt}},
\ee
where $\Delta\omega=2\pi/(t_f-t_i)$ and $\omega_k=k\Delta\omega$. First we remove the IR cutoff by performing the limit $t_f-t_i\to\infty$. One can simplify this expression for sufficiently large $t_f-t_i$ by splitting the values of $k$ into two sets, for $1\le n<cN$ and for $cN\le n<N$ where $c$ is a small, $N$-independent number. The sum over the latter set is $\ord{1/N}$ and can be neglected for large $N$, allowing the replacement, $\frac2\dt\sin\frac{\omega_k\dt}2\to\omega_n$, in the right hand side of \eq{propsum}. Another, simpler justification of such a replacement is simply noting that the sum has $-1$ overall degree of divergence therefore it converges uniformly and the summation and the limit $\dt\to0$ commute. The result is the expression
\be\label{fpropho}
\Tr_\dt[T[x(t)x(t')]]=\frac{i\hbar}m\int^{\frac\pi\dt}_{-\frac\pi\dt}\frac{d\omega}{2\pi}\frac{e^{-i\omega(t-t')}}{\omega^2-\Omega^2+i\epsilon},
\ee
for the UV regulated propagator. 

To check the canonical commutation relation we consider the function
\be
C(t,\dt)=\frac{m}t\Tr_\dt[x(t)(x(t)-x(0))-(x(t)-x(0))x(0)].
\ee
The insertion of the momentum, $p_\ell$, into a matrix element is equivalent with the presence of the multiplicative factor $m(x_\ell-x_{\ell-1})/\dt$ in the integrand of the path integral. Therefore the expectation value of the canonical commutator is
\be
\lim_{\dt\to0}C(\dt,\dt)=\la0|[x,p]|0\ra.
\ee
The factor $1/t$ in front of the trace brings the primitiv degree of divergence up to zero and the limit $\dt\to0$ non-uniform for $0\le t$. In fact, the sum
\be
C(t,\dt)=\frac{i\hbar}{\pi t}\Delta\omega\sum_{k=1}^N\frac{1-e^{-i2\pi k\frac{t}{t_f-t_i}}}{\frac4{\dt^2}\sin^2\frac{\omega_k\dt}2-\Omega^2+i\frac\epsilon{m\dt}},
\ee
gives $\lim_{t\to0}C(t,\dt)=0$ but it can safely be replaced by the integral
\be
C(t,\dt)=i\frac\hbar{\pi t}\int^{\frac\pi\dt}_{-\frac\pi\dt}d\omega\frac{1-e^{-i\omega t}}{\omega^2-\Omega^2+i\epsilon},
\ee
for a fixed $t>0$, yielding $\lim_{t\to0}C(t,0)=i\hbar$, in agreement with the canonical commutation relation. This latter implies an unbounded spectrum of the canonical variables, and it is natural that the UV cutoff must be removed before the limit $t\to0$ is performed. 

Another consequence of the canonical commutation relation and the non-uniform convergence of the Feynman propagator is the linear UV divergence of
\bea
\lim_{t'\to t}\Tr_\dt[T[\dot x(t)\dot x(t')]]&=&\frac{i\hbar}m\int^{\frac\pi\dt}_{-\frac\pi\dt}\frac{d\omega}{2\pi}\frac{\omega^2}{\omega^2-\Omega^2+i\epsilon}\nn
&=&\frac{i\hbar}{m\dt}+\frac{\hbar\Omega}{2m}+\ord{\dt},
\eea
indicating that the trajectories which dominate the path integral have diverging velocity. The scaling law $v=\ord{\dt^{-1/2}}$ implies that the dominant trajectories of the path integral are fractals with Haudorff dimension $D_H=2$ and the time integral of the action is of the \^Ito-type \cite{schulman}.

\subsection{Long time cutoff: null-space singularities and auxiliary  conditions}\label{masssings}

We consider now a harmonic oscillator whose coordinate couples linearly to a time dependent external source, the corresponding Lagrangian being $L+jx$ where $L$ is given by eq. \eq{harmosclagr}. The null space of the oscillator consists of the trajectories satisfying the homogeneous equation of motion, $(\partial_t^2+\Omega^2)x(t)=0$, and it plays a special role in the dynamics. One reason for this is that the action is degenerate within the null-space hence the trajectories from the null-space do not appear in the variational principle. We use the eigenfunctions $(\partial_t^2+\Omega^2)x_\lambda(t)=\lambda x_\lambda(t)$ as a basis set for periodic trajectories with period length $\tau$ and write the solution of the inhomogeneous equation of motion as a linear superposition,
\be\label{modedec}
x(t)=-\sum_\lambda\frac{j_\lambda}\lambda x_\lambda(t),
\ee
where $j_\lambda$ is the component of the external source in our basis. To allow the desired initial or final conditions at $t_i$ or $t_f$, respectively, we go to the limit $\tau\to\infty$ and replace the sum by an integral. The impact of the external source on the dynamics is strong at $\lambda\sim0$ and we have to exclude the null space from the discrete spectrum of the source. The Fourier integral, appearing in eq. \eq{modedec}, develops a logarithmic divergence  due to the null-space modes and needs regularization. To resolve the null space we need a long time and therefore use a regulator which modifies the Green function to 
\be\label{neargf}
D_T(t)=\int\frac{d\omega}{2\pi}\Theta\left(||\omega|-\Omega|-\frac1T\right)\frac{e^{-i\omega t}}{m(\omega^2-\Omega^2)}
\ee
yielding a solution in the form
\be\label{greensol}
x(t)=\lim_{T\to\infty}\int dtD_T(t-t')j(t').
\ee
The regulator can easily be removed with the result
\be
\lim_{T\to\infty}D_T(t)=\int\frac{d\omega}{2\pi}P\frac{e^{-i\omega t}}{m(\omega^2-\Omega^2)}
=-\frac{\sin\Omega|t|}{2m\Omega}=D^n(t),
\ee
$P$ standing for the principal value prescription. $D^n(t)$ can be called the near Green function by analogy with electrodynamics. The kind of divergences we encounter here usually shows up when the continuous and the discrete spectra overlap and their separation needs a low frequency regulator. 

Another special role of the null space can be found by noting that the near Green function is symmetric under time inversion, $D^n(t)=D^n(-t)$, hence can not accomodate initial or final conditions which break the time reversal symmetry. These conditions are usually realized by adding an appropriately chosen trajectory from the null space to the right hand side of eq. \eq{greensol}. This strategy leads to the retarded and advanced Green functions
\be\label{retadvgf}
D^{\stackrel{r}{a}}(t)=D^n(t)\pm D^f(t)
\ee
withe the far Green function
\be
D^f(t)=-\frac{\sin\Omega t}{2m\Omega}.
\ee
The procedure works because principal value integral can completely be canceled by null-space trajectories for $t>0$ or $t<0$. Such a cancelation is highly non-trivial and occurs due to the completeness of the eigenfunctions of $\partial_t^2+\Omega^2$ in the limit $t_i\to-\infty$, $t_f\to\infty$,
\be
\sum_\lambda x_\lambda(t)x_\lambda(t')\to\delta(t-t').
\ee

The construction of the solution of the initial or final condition problem is streamlined by the traditional $\epsilon$-prescription, 
\bea
\frac1{x+i\epsilon}&=&\frac{x}{x^2+\epsilon^2}-i\frac\epsilon{x^2+\epsilon^2}\nn
&=&P\frac1x-i\pi\delta(x)
\eea
which reproduces the Green function \eq{retadvgf} by the appropriate, infinitesmial shift of the poles. The first equation, written for finite $\epsilon$, indicates that the $\epsilon$-prescription is actually an IR regularization procedure. 

\begin{figure}
\centerline{\includegraphics[scale=1]{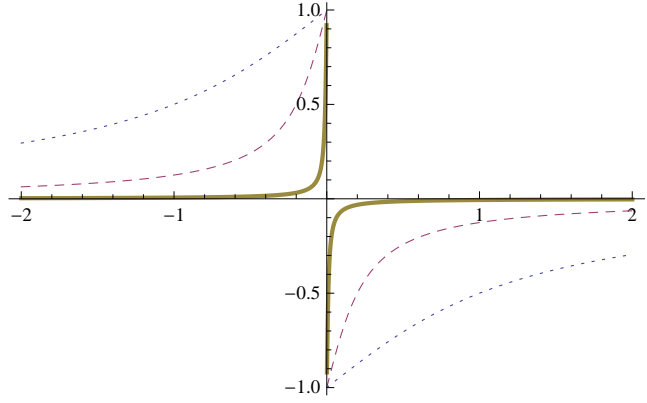}}
\caption{The function $G(\epsilon T)$ plotted against $\epsilon$ for $T=1$ (dotted line), $T=5$ (dashed line) and $T=100$ (thick line).}\label{arrf}
\end{figure}

We have now two regulators: $T$ provided by the finite time of observations, and $\epsilon$ handling the auxilary conditions. Their simultaneous presence leads to non-uniform convergence in \eq{greensol} when the retarded Green function
\be\label{regretgf}
D^r_T(t)=\int\frac{d\omega}{2\pi}\Theta\left(||\omega|-\Omega|-\frac1T\right)\frac{e^{-i\omega t}}{m[\omega^2-\Omega^2+i\sign(\omega)\epsilon]}
\ee
is used. For large $T$
\be\label{regretgffe}
D^r_T(t)=-[\Theta(t)+G(\epsilon T)]\frac{\sin\Omega t}{m\Omega}+\ord{\frac{t}T},
\ee
with
\be
G(z)=\frac1\pi\Im\ln\left(\frac{iz+1}{iz-1}\right),
\ee
cf. Fig. \ref{arrf}. The similarities of Figs. \ref{magnf} and \ref{arrf} arise from the relation $G(z)=F(z)-1\mod2$, showing that $\epsilon$ play the role of an external parameter to break the time reversal symmetry explicitely. The symmetry of the solution \eq{greensol} remains broken after the limit $\epsilon\to0$, similar to the usual spontaneous symmetry breaking generated by the auxiliary conditions via the kinetic energy.

\section{Green function of a harmonic oscillator}\label{grhos}

We calulate in this Appendix the chronon Green function of a harmonic oscillator which belongs to the trivial initial conditions $x(t_i)=\dot x(t_i)=0$.

\subsection{Discretized time}

The irregular dynamics at the initial and the end points of the trajectories require special care hence we use discretized time, $t_n=t_i+n\dt$, $n=0,\ldots,N$, with $\dt=T/N$, $T=t_f-t_i$ and write the action \eq{lholagrc} in the form
\be
S(\hx,y,\hj)=\dt\sum_{n=0}^{N-1}\sum_{\sigma=\pm1}\left\{\sigma\left[\frac{m}2\left(\frac{x^\sigma_{n+1}-x^\sigma_n}\dt\right)^2-\frac{m\Omega^2}2x^{\sigma2}_n+j^\sigma_nx^\sigma_n\right]+i\frac{m\epsilon}2x^{\sigma2}_n\right\},
\ee
where $\hx=(\hx_0,\ldots,\hx_{N-1})$, $x^\pm_0=0$, and $y=x_N$, written as
\be\label{condhoect}
S(\hx,y,\hj)=\hf\hx\hD^{-1}_0\hx+\sum_{n=0}^{N-1}\hx(y\hat A+\hj),
\ee
the trajectory $\hx_n$ being treated as an $N$-dimensional vector, 
\be\label{invpropho}
D^{-1}_{0(\sigma n)(\sigma'n)}=\delta_{\sigma\sigma'}\sigma\left[\frac{m}{\dt}(2\delta_{n,n'}-\delta_{n+1,n'}-\delta_{n-1,n'})-\dt m\Omega^2\delta_{n,n'}+i\dt\sigma m\epsilon\delta_{n,n'}\right],
\ee
and
\be
A_{\sigma n}=-\delta_{n,N-1}\frac{m\sigma}{\dt}.
\ee

To find $\hD_0$ we choose $t_i=-T$, $t_f=0$, use the linear space of functions defined at discrete time values,  $x(t_n)$, $n=0,\ldots,N$,  within the interval $[-T,0]$ with boundary conditions $x(t_0)=x(t_N)=0$,
\be\label{funcspace}
x(t_n)=\sqrt{\frac2{T}}\sum_{j=1}^Nx_j\sin\omega_jn,
\ee
with $\omega_j=j\pi/T$. The inverse of \eq{invpropho} is easy to find,
\be\label{ctpgrfndiag}
\hD_0=\begin{pmatrix}D_0&0\cr0&D_0^*\end{pmatrix},
\ee
where
\be\label{greenho}
D_0(t,t')=\frac2{Tm}\sum_{j=1}^N\frac{\sin\omega_jt\sin\omega_jt'}{\hat\omega^2_j-\Omega^2+i\epsilon}
\ee
and $\hat\omega_n=\frac2\dt\sin\pi\frac{\dt n}{2T}$.

\subsection{Effective action} 

The closing of the chronon trajectory at the final time is realized by the shift $\hx(t_n)\to\hx(t_n)+\delta_{n,N}x_f$, $x_f$ being the common final point. Since we are not interested in the dynamics of $x_f$ it can be eliminated by employing the method of the effective theory outlined in Section \ref{effacts}. First we perform the Legendre transformation $W[\hj]=S[\hx,x_f,\hj]$,
\be\label{ltreqs}
\frac{\partial S(\hx,x_f,\hj)}{\partial\hx}=-\hj,~~~~~~
\frac{\partial S(\hx,x_f,\hj)}{\partial x_f}=0.
\ee
We eliminate now $\hx$ by solving the first equation, $\hx=-\hD_0(x_f\hat A+\hj)$ and insert the solution into \eq{condhoect},
\be\label{secltrho}
S(\hx,x_f,\hj)=-\hf(x_f\hat A+\hj)\hD_0(y\hat A+\hj).
\ee
The solution of the second equation of \eq{ltreqs},
\be
y=-\frac{\hat A\hD_0\hj}{\hat A\hD_0\hat A},
\ee
is inserted into \eq{secltrho} to give
\be
W[\hj]=-\hf\hj\hD\hj,
\ee
in terms of the Green function
\be\label{ctpgrfnexp}
\hD=\hD_0-\hD_0\hat A\frac1{\hat A\hD_0\hat A}\hat A\hD_0.
\ee

\subsection{Removal of the UV and the IR cutoffs}

To remove the UV cutoff we need
\be
\frac{m^2}{\dt^2}D_0(-\dt,-\dt)=\frac{2m}{T\dt^2}\sum_{j=1}^N\frac{\sin\omega^2_j\dt}{\hat\omega^2_j-\Omega^2+i\epsilon}
\ee
which we split into a sum of $\ord{N}$ and finite pieces,
\be\label{dttdtt}
\lim_{N\to\infty}\frac{m^2}{\dt^2}D_0(-\dt,-\dt)=\frac{2m}T\sum_{j=1}^\infty\left(1-\sin^2\frac{\dt\omega_j}2\right)+\frac{2m\Omega^2}T\sum_{j=1}^\infty\frac1{\omega^2_j-\Omega^2+i\epsilon}.
\ee
The other ingredient is
\be\label{ttdt}
\lim_{\dt\to0}\frac{m}\dt D_d(t,-\dt)=-\frac2T\sum_{j=1}^\infty\frac{\omega_j\sin\omega_jt}{\omega^2_j-\Omega^2+i\epsilon}.
\ee

The CTP diagonal Green function, $\hD_0$, can be found in a manner, similar to the derivation of the Feynman propagator \eq{fpropho}, the result being
\be\label{propstal}
\lim_{\dt\to0}D_0(t,t')=\frac2{Tm}\sum_{j=1}^\infty\frac{\sin\omega_jt\sin\omega_jt'}{\omega^2_j-\Omega^2+i\epsilon}.
\ee

The removal of the IR cutoff is done by taking a large $t_f-t_i$ and approximating the sum by an integral,
\bea\label{ingred}
D_0(t,t')&=&\frac{i}{2m\Omega}[e^{(i\Omega+\frac{\epsilon}{2\Omega})(t+t')}-e^{-(i\Omega+\frac{\epsilon}{2\Omega})|t-t'|}],\nn
\frac{m^2}{\dt^2}D_{0-\dt,-\dt}&=&\frac{m}\dt-im\Omega,\nn
\frac{m}\dt D_0(t,-\dt)&=&-e^{(i\Omega+\frac{\epsilon}{2\Omega})t},
\eea
assuming the inequality $\Delta\omega=\pi/T\ll\epsilon/\Omega$ to assure the applicability of the contour integral method in the calculation. The expressions \eq{ingred} yield the Green function
\be\label{fgrfncthoc}
\hD(t,t')=-\frac{i}{2m\Omega}\left[\begin{pmatrix}e^{-(i\Omega+\frac{\epsilon}{2\Omega})|t-t'|}&0\cr0&-e^{(i\Omega+\frac{\epsilon}{2\Omega})|t-t'|}\end{pmatrix}+e^{\frac{\epsilon}{2\Omega}(t+t')}\begin{pmatrix}0&e^{i\Omega(t-t')}\cr e^{-i\Omega(t-t')}&0\end{pmatrix}\right],
\ee
which becomes in the limit $t_i\to-\infty$, $t_f\to\infty$
\be\label{fgrfnctho}
\hD(t,t')=\frac{i}{2m\Omega}\begin{pmatrix}-e^{-i\Omega|t-t'|}&-e^{i\Omega(t-t')}\cr e^{-	i\Omega(t-t')}&e^{i\Omega|t-t'|}\end{pmatrix},
\ee
ie.
\be\label{dnfi}
D^n=-\frac{\sin\Omega|t|}{2m\Omega},~~~
D^f=-\frac{\sin\Omega t}{2m\Omega},~~~
D^i=-\frac{\cos\Omega t}{2m\Omega},
\ee
as long as the inequality, $t_f-t\ll\Omega/\epsilon$ is observed to prevent the loss of the final condition \eq{ctpconstr}. Note that the Green function corresponds to the initial conditions $x_i=\dot x_i=0$ within our functional space.

\subsection{Quantum oscillator}\label{qgfn}

The generator functional \eq{ctpgf} for the harmonic oscillator for $\rho_i=|0\ra\la0|$,
\be
W[\hj]=-\hf\hat{\bar j}\hD\hj,
\ee
reproduces the classical Green function, written in this case as
\be
i\hD(t,t')=\begin{pmatrix}\la T[x(t)x(t')]\ra&\la x(t')x(t)\ra\cr\la x(t)x(t')\ra&\la T[x(t')x(t)]\ra^*\end{pmatrix}.
\ee
In particular, $D^{++}$ is the Feynman propagator and $D^{-+}$ gives the Wightman function.

\end{document}